\documentstyle[amssymb,multicol,amsmath]{article}
\newtheorem{Proposition}{Proposition}

\newcommand{\bp}{\begin{Proposition}}
\newcommand{\ep}{\end{Proposition}}
\newcommand{\bi}[1]{\vspace{-3mm} \bibitem{#1}}

\begin{document}

\begin{center}
{\Large \bf Open n-Qubit System as a Quantum Computer \\
with Four-Valued Logic}
\vskip 7 mm
{\Large \bf Vasily E. Tarasov } \\
\vskip 5mm
{\it Skobeltsyn Institute of Nuclear Physics,\\
Moscow State University, Moscow 119899, Russia}\\
\vskip 3 mm
{E-mail: TARASOV@THEORY.SINP.MSU.RU}\\
\vskip 21 mm

{\large \bf Preprint of Skobeltsyn Institute of Nuclear Physics,\\
Moscow State University, Moscow \\
\vskip 4mm
SINP MSU 2001-31/671 \ \ \ 20 August 2001}

\end{center}

\vskip 5 mm
\begin{abstract}
{\large In this paper we generalize the usual model of quantum
computer to a model in which the state is an operator of density
matrix and the gates are general superoperators (quantum
operations), not necessarily unitary. A mixed state (operator of
density matrix) of  n two-level quantum system (open or closed
n-qubit system) is considered as an element of $4^{n}$-dimensional
operator Hilbert space (Liouville space). It allows to use quantum
computer (circuit) model with 4-valued logic. The gates of this
model are general superoperators which act on n-ququats state.
Ququat is quantum state in a 4-dimensional (operator) Hilbert
space. Unitary two-valued logic gates and quantum operations for
n-qubit open system
are considered as four-valued logic gates acting on n-ququats.
We discuss properties of quantum 4-valued logic gates.
In the paper we study universality for 
quantum four-valued logic gates.
}
\end{abstract}

PACS {3.67.Lx; 03.65-w; 3.65.Bz}



\twocolumn


\section{Introduction}


Usual models for quantum computer use closed n-qubit systems and
deal only with unitary gates
on pure states. In these models it is difficult or impossible
to deal formally with measurements, dissipation, decoherence and noise.
It turns out, that the restriction to pure states and unitary gates
is unnecessary \cite{AKN}.

One can describe an open system starting from a closed system
if the open system is a part of the closed system.
However, situations can arise where it is difficult or impossible
to find a closed system comprising the given open system.
This would render theory of dissipative and open systems
a fundamental generalization of quantum mechanics \cite{Dav}.
Understanding dynamics of open systems is important for studying
quantum noise processes \cite{Gar,Schu,SN}, quantum error
correction \cite{Sh,St2,CS,Pr,BDSW}, decoherence effects
\cite{Unr,PSE,BBSS,MPP,Om,LCW,BLW,Zan,ZR} in quantum computations
and to perform simulations of open quantum systems
\cite{Ter,TDV,Bac,LV,AbL,Zal}.

In this paper we generalize the usual model of quantum computer
to a model in which the state is a density matrix operator
and the gates are general superoperators (quantum operations),
not necessarily unitary.
Pure state of n two-level closed quantum systems is an element
of $2^{n}$-dimensional Hilbert space and it allows to realize
quantum computer model with 2-valued logic. The gates of this
computer model are unitary operators act on a such state.
In general case, mixed state (operator of density matrix)
of  n two-level quantum systems is an element of
$4^{n}$-dimensional operator Hilbert space (Liouville space).
It allows to use quantum computer model with 4-valued logic. The
gates of this model are general superoperators (quantum
operations) which act on general n-ququats state. Ququat
\cite{BPT} is quantum state in a 4-dimensional (operator)
Hilbert space. Unitary gates and quantum operations for quantum
two-valued logic computer can be considered as four-valued logic
gates of new model. In the paper we consider universality for
general quantum 4-valued logic gates acting on ququats.

In Sections 2, 3 and 5, the physical and mathematical background
(pure and mixed states, Liouville space and superoperators,
evolution equations for closed and open quantum) are considered.
In Section 4, we introduce generalized computational basis and
generalized computational states for $4^{n}$-dimensional operator
Hilbert space (Liouville space). In the Section 6, we study some
properties of general four-valued logic gates. Unitary gates and
quantum operations of two-valued logic computer are considered as
four-valued logic gates. In Section 7, we introduce a four-valued
classical logic formalism. In Section 8, we realize classical
4-valued logic gates by quantum gates. In Section 9, we consider a
universal set of quantum 4-valued logic gates. In Section 10,
quantum four-valued logic gates of order (n,m) as a map from
density matrix operator on n-ququats to density matrix operator on
m-ququats are discussed.



\section{Quantum state and qubit}


\subsection{Pure states}

A quantum system in a pure state is described by unit vector in a
Hilbert space ${\cal H}$. In the Dirac notation a pure state is
denoted by $|\Psi>$. The Hilbert space ${\cal H}$ is a linear
space with an inner product. The inner product  for $ |\Psi_{1}
>$, $|\Psi_{2}> \in {\cal H}$ is denoted by $<\Psi_{1}|\Psi_{2}>
$. A quantum bit or qubit, the fundamental concept of quantum
computations, is a two-state quantum system. The two basis states
labeled $|0>$ and $|1>$, are orthogonal unit vectors, i.e.
\[ <k|l>=\delta_{kl}, \]
where $k,l\in \{0,1\}$.
The Hilbert space of qubit is ${\cal H}_{2}={\mathbb{C}}^2$.
The quantum system which corresponds to a quantum computer
(quantum circuits) consists of n quantum two-state particles.
The Hilbert space ${\cal H}^{(n)}$ of such a system is a tensor product
of n Hilbert spaces ${\cal H}_{2}$ of one two-state particle: \
${\cal H}^{(n)}={\cal H}_{2}
\otimes {\cal H}_{2} \otimes ... \otimes {\cal H}_{2}$.
The space ${\cal H}^{(n)}$ is a $N=2^{n}$ dimensional complex linear
space. Let us choose a basis for ${\cal H}^{(n)}$
which is consists of the $N=2^{n}$ orthonormal states
$|k>$, where k is in binary representation.
The state $|k>$ is a tensor product of the states $|k_i>$ in ${\cal H}^{(n)}$:
\[ |k>=|k_{1}>\otimes |k_{2}> \otimes ... \otimes |k_{n}>=
|k_{1}k_{2}...k_{n}>  \ , \]
where $k_{i} \in \{0,1\}$ and $i=1,2,...,n$.
This basis is usually called computational basis
which has $2^{n}$ elements.
A pure state $|\Psi(t)> \in {\cal H}^{(n)}$ is generally a superposition
of the basis states
\begin{equation} \label{Psi} |\Psi(t)>=\sum^{N-1}_{k=0}a_{k}(t)|k> \ ,
\end{equation}
with $N=2^{n}$ and $\sum^{N-1}_{k=0} |a_{k}(t)|^{2}=1$.
The inner product between  $|\Psi>$ and $|\Psi'>$ is denoted by
$<\Psi|\Psi'>$ and
\[ <\Psi|\Psi'>=\sum^{N-1}_{k=0} a^{*}_{k} {a'}_{k}  \ . \]

\subsection{Mixed states}

In general, a quantum system is not in a pure state. Open quantum
systems are not really isolated from the rest of the universe, so
it does not have a well defined pure state. Landau and von Neumann
introduced a mixed state and a density matrix into quantum theory.
A density matrix is a Hermitian ($\rho^{\dagger}=\rho$), positive
($\rho >0$) operator on ${\cal H}^{(n)}$ with unit trace ($Tr \rho =1$).
Pure states can be characterized by idempotent condition $\rho^{2}=\rho$. 
A pure state (\ref{Psi}) is represented by the operator
$\rho=|\Psi><\Psi|$.

One can represent an arbitrary density matrix operator $\rho(t)$
for $n$-qubits in terms of tensor products of Pauli
matrices $\sigma_{\mu}$:
\begin{equation}
\label{rhosigma} \rho(t)=\frac{1}{2^{n}} \sum_{\mu_1 ... \mu_n}
P_{\mu_1 ... \mu_n}(t)
\sigma_{\mu_1} \otimes ... \otimes \sigma_{\mu_n} \ . \end{equation}
where each $\mu_{i} \in \{0,1,2,3\}$ and $i=1,...,n$.
Here $\sigma_{\mu}$ are Pauli matrices
$$ \label{sigma}
\sigma_{1}=\left(
\begin{array}{cc}
0&1\\
1&0\\
\end{array}
\right), \ \ \
\sigma_{2}=\left(
\begin{array}{cc}
0&-i\\
i&0\\
\end{array}
\right), $$
$$\sigma_{3}=\left(
\begin{array}{cc}
1&0\\
0&-1\\
\end{array}
\right), \ \ \
\sigma_{0}=I=\left(
\begin{array}{cc}
1&0\\
0&1\\
\end{array}
\right).
$$
The real expansion coefficients $P_{\mu_1 ... \mu_n}(t)$ are given by
\[ P_{\mu_1 ... \mu_n}(t)=Tr( \sigma_{\mu_1} \otimes ...
\otimes \sigma_{\mu_n} \rho(t)). \] Normalization ($Tr \rho=1$)
requires that $P_{0...0}(t)=1$. Since the eigenvalues of the Pauli
matrices are $\pm 1$, the expansion coefficients satisfy
$|P_{\mu_1...\mu_n}(t)|\le 1$. Let us rewrite (\ref{rhosigma}) in
the form:
\[ \rho(t)=
\frac{1}{2^n}\sum^{N-1}_{\mu=0} \sigma_{\mu} P_{\mu}(t), \]
where $\sigma_{\mu}=\sigma_{\mu_1} \otimes ... \otimes \sigma_{\mu_n}$,
$\mu=(\mu_1...\mu_n)$ and $N=4^{n}$.

An arbitrary general one-qubit state $\rho(t)$ can be represented as
\[ \rho(t)=\frac{1}{2} \sum^{3}_{\mu=0} \sigma_{\mu} P_{\mu}(t), \]
where $P_{\mu}(t)=Tr(\sigma_{\mu} \rho(t))$ and $P_0(t)=1$.
The pure state can be identified with Bloch sphere
\[ P^2_1(t)+P^2_2(t)+P^2_3(t)=1. \]
The mixed state can be identified with close ball
\[ P^2_1(t)+P^2_2(t)+P^2_3(t) \le 1. \]

Not all linear combinations of quantum states $\rho_j(t)$ are states.
The operator
\[ \rho(t)=\sum_j \lambda_j \rho_j (t) \]
is a state iff  \ $\sum_j \lambda_j=1$.


\section{Liouville space and superoperators}


For the concept of Liouville space and superoperators see
\cite{Cra}-\cite{kn2}.

\subsection{Liouville space}

The space of linear operators acting on a
$N=2^{n}$-dimensional Hilbert space ${\cal H}^{(n)}$
is a $N^{2}=4^{n}$-dimensional complex linear space
$\overline {\cal H}^{(n)}$.
We denote an element $A$ of $\overline{\cal H}^{(n)}$
by a ket-vector $|A)$. The inner product of two elements $|A)$ and
$|B)$ of $\overline{\cal H}^{(n)}$ is defined as
\begin{equation} \label{inner} (A|B)=Tr(A^{\dagger} B) \ . \end{equation}
The norm $\|A\|=\sqrt{(A|A)}$ is the Hilbert-Schmidt norm of
operator $A$. A new Hilbert space $\overline{\cal H}$ with scalar
product (\ref{inner}) is called Liouville space attached to ${\cal
H}$ or the associated Hilbert space, or Hilbert-Schmidt space
\cite{Cra}-\cite{kn2}.

Let $\{|k>\}$ be an orthonormal basis of ${\cal H}^{(n)}$:
\[ <k|k'>=\delta_{kk'} \ , \quad \sum^{N-1}_{k=0}|k><k|=I. \]
Then $|k,l)=||k><l|)$
is an orthonormal basis of the Liouville space $\overline{\cal H}^{(n)}$:
\begin{equation} \label{onb} (k,l|k',l')=\delta_{kk'}\delta_{ll'} \ , \quad
\sum^{N-1}_{k=0} \sum^{N-1}_{l=0}|k,l)(k,l|=\hat I  \ , \end{equation}
where $N=2^n$. This operator basis has $4^{n}$ elements.
Note that $|k,l) \not= |kl>=|k>\otimes |l>$ and
\begin{equation} \label{k,l}
|k,l)=|k_{1},l_{1})\otimes |k_{2},l_{2}) \otimes ...
\otimes |k_{n},l_{n}) \ , \end{equation}
where $k_{i},l_{i} \in \{0,1\}$, $i=1,...,n$ and
\[ |k_{i},l_{i})\otimes |k_{j},l_{j})=
| \ |k_{i}>\otimes  |k_{j}>, <l_{i}| \otimes <l_{j}| \ ). \]
For an arbitrary element $|A)$ of $\overline{\cal H}^{(n)}$ we have
\begin{equation} \label{|A)}
|A)=\sum^{N-1}_{k=0}\sum^{N-1}_{l=0} |k,l)(k,l|A) \end{equation}
with
\[ (k,l|A)=
Tr( |l><k| A )=<k|A|l>=A_{kl}. \]
An operator $\rho(t)$ of density matrix for n-qubits can be considered
as an element $|\rho(t))$ of the space $\overline{\cal H}^{(n)}$.
From (\ref{|A)}) we get
\begin{equation} \label{|rho)} |\rho(t))=\sum^{N-1}_{k=0}
\sum^{N-1}_{l=0} |k,l)(k,l|\rho(t)) \ , \end{equation}
where $N=2^n$ and
\[ \sum^{N-1}_{k=0} (k,k|\rho(t))=1. \]


\subsection{Superoperators}

Operators, which act on $\overline{\cal H}$, are called superoperators
and we denote them in general by the hat.


For an arbitrary superoperator $\hat \Lambda$ on
$\overline{\cal H}$, which is defined by
\[ \hat \Lambda|A)=|\Lambda(A) ), \]
we have
\[ (k,l|\hat \Lambda|A)=\sum^{N-1}_{k'=0}\sum^{N-1}_{l'=0}
(k,l|\hat \Lambda|k',l') (k',l'|A)=\]
\[ =\sum^{N-1}_{k'=0}\sum^{N-1}_{l'=0}
\Lambda_{klk'l'}A_{k'l'}, \]
where $N=2^n$.

Let $A$ be a linear operator in Hilbert space.
Then the superoperators $\hat L_{A}$ and $\hat R_{A}$ will be defined by
\[ \hat L_{A}|B)=|AB) \ , \quad \hat R_{A}|B)=|BA). \]
In the basis $|k,l)$ we have
\[ (k,l|\hat L_{A}|B)=\sum^{N-1}_{k'=0}
\sum^{N-1}_{l'=0}
(k,l|\hat L_{A}|k',l')(k',l'|B)=\]
\[ =\sum^{N-1}_{k'=0}\sum^{N-1}_{l'=0}(\hat L_{A})_{klk'l'} <k'|B|l'>, \]
Note that
\[ (k,l|AB)=<k|AB|l>=\]
\[ =\sum^{N-1}_{k'=0}\sum^{N-1}_{l'=0}
<k|A|k'><k'|B|l'><l'|l>. \]
Finally, we obtain
\[ (\hat L_{A})_{klk'l'} =<k|A|k'><l'|l>=A_{kk'}\delta_{ll'}. \]
The superoperator $\hat P=|A)(B|$ is defined by
\[ \hat P|C)=|A)(B|C)=|A)Tr(B^{\dagger}C). \]

A superoperator $\hat{\cal E}^{\dagger}$ is called the adjoint
superoperator for $\hat{\cal E}$ if
\[ (\hat{\cal E}^{\dagger}(A)|B)=(A|\hat{\cal E}(B)) \]
for all $|A)$ and $|B)$ from $\overline{\cal H}$.
For example, if $\hat{\cal E}=\hat L_{A}\hat R_{B}$, then
$\hat{\cal E}^{\dagger}=\hat L_{A^{\dagger}}\hat R_{B^{\dagger}}$.
If $\hat{\cal E}=\hat L_{A}$, then
$\hat{\cal E}^{\dagger}=\hat L_{A^{\dagger}}$.
If $\hat{\cal E}=\hat L_{A}\hat R_{A^{\dagger}}$, then
$\hat{\cal E}^{\dagger}=\hat L_{A^{\dagger}}\hat R_{A}$.

A superoperator $\hat{\cal E}$ is called unital if $\hat{\cal E}|I)=|I)$.


\section{Generalized computational basis and ququats}


Let us introduce generalized computational basis and generalized
computational  states for $4^{n}$-dimensional operator Hilbert space
(Liouville space).

\subsection{Pauli representation}

Pauli matrices (\ref{sigma}) can be considered as a basis in
operator space. Let us write the Pauli matrices (\ref{sigma}) in
the form
\[ \sigma_{1}=|0><1|+|1><0|=|0,1)+|1,0), \]
\[ \sigma_{2}=-i|0><1|+i|1><0|=-i(|0,1)-|1,0) ), \]
\[ \sigma_{3}=|0><0|-|1><1|=|0,0)-|1,1), \]
\[ \sigma_{0}=I=|0><0|+|1><1|=|0,0)+|1,1). \]
Let us use the formulas
\[ |0,0)=\frac{1}{2}(|\sigma_{0})+|\sigma_{3})) \ , \quad
|1,1)=\frac{1}{2}(|\sigma_{0})-|\sigma_{3})), \]
\[ |0,1)=\frac{1}{2}(|\sigma_{1})+i|\sigma_{2})) \ , \quad
|1,0)=\frac{1}{2}(|\sigma_{1})-i|\sigma_{2})). \]
It allows to rewrite operator basis
\[ |k,l)=|k_{1},l_{1})\otimes |k_{2},l_{2}) \otimes ...
\otimes |k_{n},l_{n}) \]
by complete basis operators
\[ |\sigma_{\mu})=|\sigma_{\mu_1} \otimes \sigma_{\mu_2} \otimes ...
\otimes \sigma_{\mu_n}), \]
where $\mu_i=2k_{i}+l_{i}$,
i.e. $\mu_{i} \in \{0,1,2,3\}$ and $i=1,...,n$.
The basis $|\sigma_{\mu})$ is orthogonal
\[ (\sigma_{\mu}|\sigma_{\mu '})=2^{n} \delta_{\mu \mu '} \]
and complete operator basis
\[ \frac{1}{2^{n}} \sum^{N-1}_{\mu}
|\sigma_{\mu})(\sigma_{\mu}|=\hat I. \]
For an arbitrary element $|A)$ of $\overline{\cal H}^{(n)}$ we have
Pauli representation by
\[ |A)=\frac{1}{2^{n}}\sum^{N-1}_{\mu=0}
|\sigma_{\mu})(\sigma_{\mu}|A) \]
with the complex coefficients
\[ (\sigma_{\mu}|A)=Tr( \sigma_{\mu} A ). \]
We can rewrite formulas (\ref{rhosigma}) using the complete
operator basis $|\sigma_{\mu})$ in Liouville space $\overline{\cal
H}^{(n)}$:
\[ |\rho(t))=\frac{1}{2^n}\sum^{N-1}_{\mu=0}
|\sigma_{\mu})(\sigma_{\mu}|\rho(t)), \]
where $\sigma_{\mu}=\sigma_{\mu_1} \otimes ... \otimes \sigma_{\mu_n}$,
\ $\mu=(\mu_1...\mu_n)$ and $(\sigma_{\mu}|\rho(t))=P_{\mu}(t)$.

The density matrix operator $\rho(t)$ is a self-adjoint operator with
unit trace. It follows that
\[ P^{*}_{\mu}(t)=P_{\mu}(t) \ , \quad P_{0}(t)=(\sigma_{0}|\rho(t))=1. \]
In general case,
\[ \frac{1}{2^n} \sum^{N-1}_{\mu=0} P^{2}_{\mu}(t)=
(\rho(t)|\rho(t))=Tr\rho^{2}(t) \le 1. \]
Note that Schwarz inequality
\[ |(A|B)|^{2} \le (A|A)(B|B)  \]
leads to
\[ |(I|\rho(t))|^{2} \le (I|I)(\rho(t)|\rho(t)), \]
\[ 1=|Tr\rho(t)|^{2} \le
2^{n}(\rho(t)|\rho(t))=\sum^{N-1}_{\mu=0} P^{2}_{\mu}(t), \]
i.e.
\[ \frac{1}{\sqrt{2^{n}}} \le Tr( \rho^{2}(t)) \le 1 \quad or \quad
1 \le \sum^{N-1}_{\mu=0} P^{2}_{\mu}(t) \le 2^{n}. \]

An arbitrary general one-qubit state $\rho(t)$ can be represented
in Liouville space $\overline{\cal H}_2$ as
\[ |\rho(t))=\frac{1}{2} \sum^{3}_{\mu=0} |\sigma_{\mu}) P_{\mu}(t), \]
where $P_{\mu}=(\sigma_{\mu} |\rho)$, $P_0=1$ and
$P^2_1(t)+P^2_2(t)+P^2_3(t) \le 1$.
Note that the basis
$|\sigma_{\mu})$ is orthogonal, but is not orthonornal.

\subsection{Generalized computational basis}

Let us define the orthonormal basis $|\mu)$ of Liouville space
$\overline{\cal H}^{(n)}$.
In general case, the state  $\rho(t)$ of the n-qubits system
is an element of Hilbert space $\overline{\cal H}^{(n)}$.
The basis for $\overline{\cal H}^{(n)}$ consists of the
$N^2=2^{2n}=4^n$ orthonormal basis elements denoted by $|\mu)$.

\noindent {\bf Definition}
{\it A basis of Liouville space $\overline{\cal H}^{(n)}$ is defined by
\[ |\mu)=|\mu_1...\mu_n)=\frac{1}{\sqrt{2^{n}}}|\sigma_{\mu})=
\frac{1}{\sqrt{2^{n}}}|\sigma_{\mu_1}
\otimes  ... \otimes \sigma_{\mu_n}), \]
where each $\mu_i \in \{0,1,2,3\}$ and
\[ (\mu|\mu')=\delta_{\mu \mu'} \ , \quad
\sum^{N-1}_{\mu=0} |\mu)(\mu|=\hat I, \]
is called a {\bf generalized computational basis}.}

Here $\mu$ is  4-valued representation of
\[ \mu=\mu_1 4^{n-1}+...+\mu_{n-1}4+\mu_n \ .\]

\noindent
{\bf Example.}
In general case, one-qubit state $\rho(t)$ of open quantum system is
\[ |\rho)=|0)\frac{1}{\sqrt{2}}+|1)\rho_{1}+
|2)\rho_{2}+|3) \rho_{3}, \]
where {\it four} orthonormal basis elements are
\[ |0)=\frac{1}{\sqrt{2}} |\sigma_0)=\frac{1}{\sqrt{2}}|I) \ , \quad
|1)=\frac{1}{\sqrt{2}} |\sigma_1), \]
\[ |2)=\frac{1}{\sqrt{2}} |\sigma_2) \ , \quad
|3)=\frac{1}{\sqrt{2}} |\sigma_3). \]
{\bf Example.}
Two-qubit state $\rho(t)$ is an element of 16-dimensional
Hilbert space with the orthonormal basis
\[ |00)=\frac{1}{2}|I\otimes I) \ , \quad
|0k)=\frac{1}{2} |I\otimes\sigma_k), \]
\[ |k0)=\frac{1}{2} |\sigma_k\otimes I) \ , \quad
|kl)=\frac{1}{2} |\sigma_{k} \otimes \sigma_l), \]
where $k, l \in \{1,2,3\}$.

The usual computational basis $\{|k>\}$ is not a basis of general
state $\rho(t)$ which has a time dependence. In general case, a
pure state evolves to mixed state.

Pure state of n two-level closed quantum systems is an element
of $2^{n}$-dimensional functional Hilbert space ${\cal H}^{(n)}$.
It leads to quantum computer model with 2-valued logic.
{\it In general case, the mixed state $\rho(t)$ of $n$ two-level
(open or closed) quantum system is an element of $4^{n}$-dimensional
operator Hilbert space $\overline{\cal H}^{(n)}$ (Liouville space).
It leads to 4-valued logic model for quantum computer.}

The state of the quantum computation at any point time is a superposition
of basis elements
\[ |\rho(t))=\sum^{N-1}_{\mu=0} |\mu)\rho_{\mu}(t), \]
where $\rho_{\mu}(t)$ are real numbers (functions)
satisfying normalized condition $\rho_0(t)=1/\sqrt{2^{n}}$, i.e.
\[ \sqrt{2^n}(0|\rho(t))=Tr(\rho(t))=1. \]
Any state $|\rho(t))$  for basis element
$|0...0)$ has $P_{0...0}=1$ in all cases.

\subsection{Generalized computational states}

Generalized computational
basis elements $|\mu)$ are not quantum states for
$\mu\not=0$. It follows from normalized condition
$(0|\rho(t))=1/\sqrt{2}$.
The general quantum state in Pauli representation (\ref{rhosigma})
has the form
\[ |\rho(t))=\frac{1}{2^{n}}\sum^{N-1}_{\mu=0}
|\sigma_{\mu})P_{\mu}(t), \]
where $P_{0}(t)=1$ in all cases.
Let us define simple computational quantum states.

\noindent {\bf Definition}
{\it A quantum states in Liouville space defined by
\[ |\mu]=\frac{1}{2^{n}}\Bigl(|\sigma_{0})+
|\sigma_{\mu}) (1-\delta_{\mu 0}) \Bigr) \]
or
\[ |\mu]=\frac{1}{\sqrt{2^{n}}}\Bigl(|0)+
|\mu)(1-\delta_{\mu 0}) \Bigr). \]
is called {\bf generalized computational states}.}

Note that all states $|\mu]$, where $\mu \not=0$, are pure states,
since $[\mu|\mu]=1$. The state $|0]$ is maximally mixed state.
The states $|\mu]$ are elements of Liouville space
$\overline{\cal H}^{(n)}$.

Quantum state in a 4-dimensional Hilbert space is usually called
ququat or qu-quart\cite{BPT} or qudit \cite{Run,CBKG,DW,BB,CMP} with $D=4$.
Usually ququat is considered as 4-level quantum system. We
consider ququat as general quantum state in a 4-dimensional
operator Hilbert space.

\noindent {\bf Definition}
{\it A quantum state in 4-dimensional operator Hilbert space
(Liouville space)
$\overline{\cal H}^{(1)}$ associated with single qubit of
${\cal H}^{(1)}={\cal H}_2$ is called {\bf single ququat}.
A quantum state in $4^n$-dimensional Liouville space
$\overline{\cal H}^{(n)}$ associated with n-qubits system
is called {\bf n-ququats}.}


\noindent
{\bf Example.}
For the single ququat the states $|\mu]$ are
\[ |0]=\frac{1}{2}|\sigma_0)  \ , \quad
|k]=\frac{1}{2}\Bigl(|\sigma_{0})+|\sigma_{k})\Bigr)\ , \]
or
\[ |0]=\frac{1}{\sqrt{2}}|0)  \ , \quad
|k]=\frac{1}{\sqrt{2}}\Bigl(|0)+|k)\Bigr). \]

It is convenient to use matrices for quantum states. In {\bf
matrix representation} the single ququat computational basis
$|\mu)$ and computational states $|\mu]$ can be represented by

$$|0)=\left(
\begin{array}{c}
1\\
0\\
0\\
0
\end{array}
\right),
|1)=\left(
\begin{array}{c}
0\\
1\\
0\\
0
\end{array}
\right),
|2)=\left(
\begin{array}{c}
0\\
0\\
1\\
0
\end{array}
\right),
|3)=\left(
\begin{array}{c}
0\\
0\\
0\\

1
\end{array}
\right).
$$

In this representation single qubit generalized computational
states $|\mu]$ is represented by
$$|0]=\frac{1}{\sqrt{2}}\left(
\begin{array}{c}
1\\
0\\
0\\
0
\end{array}
\right),
\quad
|1]=\frac{1}{\sqrt{2}}\left(
\begin{array}{c}
1\\
1\\
0\\
0
\end{array}
\right),$$
$$|2]=\frac{1}{\sqrt{2}}\left(
\begin{array}{c}
1\\
0\\
1\\
0
\end{array}
\right),
\quad
|3]=\frac{1}{\sqrt{2}}\left(
\begin{array}{c}
1\\
0\\
0\\
1
\end{array}
\right).
$$
A general single ququat quantum state
$|\rho)=\sum^{N-1}_{\mu=0}|\mu)\rho_{\mu}$ is represented
$$|\rho)=\left(
\begin{array}{c}
\rho_{0}\\
\rho_{1}\\
\rho_{2}\\
\rho_{3}
\end{array}
\right),
$$
where $\rho_{0}=1/\sqrt{2}$ and
$\rho^{2}_{1}+\rho^{2}_{2}+\rho^{2}_{3} \le \sqrt{2}$.

We can use the other matrix representation for the states
$|\rho]$ which has no the coefficient $1/\sqrt{2^{n}}$.
In this representation single qubit generalized computational
states $|\mu]$ is represented by
$$|0]=\left[
\begin{array}{c}
1\\
0\\
0\\
0
\end{array}
\right],\
|1]=\left[
\begin{array}{c}
1\\
1\\
0\\
0
\end{array}
\right],\
|2]=\left[
\begin{array}{c}
1\\
0\\
1\\
0
\end{array}
\right],\
|3]=\left[
\begin{array}{c}
1\\
0\\
0\\
1
\end{array}
\right].
$$
A general single ququat quantum state
$$|\rho]=\left[
\begin{array}{c}
1\\
P_{1}\\
P_{2}\\
P_{3}
\end{array}
\right],
$$
where $P^{2}_{1}+P^{2}_{2}+P^{2}_{3} \le 1$, is a superposition
of generalized computational states
\[ |\rho]=|0](1-P_1-P_2-P_3)+|1] P_1+|2]P_2+|3] P_3. \]

Note that density matrix operator $\rho$ as an element of
Liouville space is represented by $|\rho)$ and $|\rho]$.
We use different brackets only to emphasize the different matrix
representations connected by coefficient $1/\sqrt{2^{n}}$.
This coefficient can be neglected under the consideration of
the quantum 4-valued logic gates.


\section{Evolution equations and quantum operations}


In this section I review the description of open quantum systems
dynamics in terms of evolution equations and quantum operations.

\subsection{Evolution equation for pure state of closed systems}

Let $H$ be the Hamilton operator, then in the Schroedinger picture
the equation of motion for the pure state $|\Psi(t)>$
of closed system is given by the Schroedinger equation
\begin{equation} \label{Schr}
\frac{d}{dt}|\Psi(t)>=-i H |\Psi(t)> \ .
\end{equation}
The change in the state $|\Psi(t)>$ of a closed quantum system
between two fixed times $t$ and $t_{0}$
is described by a unitary operator $U(t,t_{0})$ which depends
on those times
\[ |\Psi(t)>=U(t,t_{0})|\Psi(t_{0})>. \]
If the Hamilton operator $H$ has no time dependence,
then the unitary operator $U(t,t_{0})$ has the form
\[ U(t,t_{0})=exp\{-i(t-t_{0})H\}. \]
In general case, the unitary operator $U(t,t_{0})$ is defined by
\[ \frac{d}{dt}U(t,t_{0})=-iHU(t,t_{0}) \ , \quad U(t_{0},t_{0})=I. \]
A pure state $|\Psi> \in {\cal H}^{(n)}$ of closed n-qubits system
is generally a superposition of the orthonormal basis states $|k>$
\[ |\Psi(t)>=\sum_{k}a_{k}(t)|k>. \]
Let the Hamilton operator $H$ on the space ${\cal H}^{(n)}$
can be written in the form
\[ H=\sum_{l,m} H_{lm} |l><m|. \]
Then
equation (\ref{Schr}) can be given in the form
\[ \frac{d}{dt} a_k(t)=-i \sum_{l} H_{kl} a_l(t). \]

\subsection{Evolution equation for mixed state of closed system}

Let $H$ be the Hamiltonian, then in the Schroedinger picture the
evolution equation for the mixed state $\rho(t)$ of closed system
is given by the von Neumann equation
\begin{equation} \label{Neu}
\frac{d}{dt}\rho(t)=-i[H,\rho(t)] \ .
\end{equation}
This equation can be rewritten by
\[ \frac{d}{dt}|\rho(t) )=\hat \Lambda |\rho(t)), \]
where the Liouville superoperator $\hat \Lambda$ is given by
\[ \hat \Lambda=-i(\hat L_{H}-\hat R_{H}). \]
A change of pure and mixed states of closed (Hamiltonian) quantum system
is the unitary evolution. The final state $\rho(t)$ of the system
is related to the initial state $\rho(t=t_0)=\rho$ by unitary
transformation $U=U(t,t_{0})$:
\begin{equation} \label{UU} \rho \ \rightarrow \
\rho(t)={\cal U}_t(\rho)=U \rho U^{\dagger} \ , \end{equation}
where $UU^\dagger=I$. The superoperator $\hat{\cal U}$ is written in the
form $\hat{\cal U}=\hat L_U \hat R_U$: \
$|\rho(t))=\hat{\cal U}_{t}|\rho)$.

\subsection{Evolution equation for mixed state of open system}

A classification of norm continuous (or, equivalently, with
bounded generators) dynamical semigroup \cite{AL} of the Bahach
space of trace-class operators on ${\cal H}$, has been given by
Lindblad (\cite{Lind1}). The general form of the generator $\hat
\Lambda$ of such a semigroup is the following
\begin{equation} \label{L1} \hat \Lambda \rho=-i[H,\rho]+\Phi(\rho) -
 \Phi(I) \circ \rho \ , \end{equation}
where \[ \Phi(B)=\sum_j V_j B V^{\dagger}_j \ ,
\quad A \circ B=\frac{1}{2}(AB+BA). \]
Here $H$ is a bounded self-adjoint Hamilton operator,
$\{V_j\}$ is a sequence of bounded operators ,
$\Phi(I)$ is a bounded operator.
The evolution equation has the form
\begin{equation} \label{L2}
\frac{d}{dt} \rho(t)=-i[H,\rho(t)]+\sum_j \Bigl( V_j \rho(t) V^{\dagger}_j
- \rho(t) \circ (V_j V^{\dagger}_j) \Bigr) \ . \end{equation}
For the proofs of (\ref{L1}) and (\ref{L2}), we refer to \cite{Lind1,AL}.
Using equation (\ref{L2}) evolution equation for the mixes state
$|\rho (t))$ can be written by
\[ \frac{d}{dt}|\rho(t) )=\hat \Lambda |\rho(t)), \]
where the Liouville superoperator $\hat \Lambda$ is given by
\[ \hat \Lambda=-i(\hat L_{H}-\hat R_{H})+
\frac{1}{2}\sum_{j}\Bigl( 2\hat L_{V_{j}} \hat R_{V^{\dagger}_{j}}-
\hat L_{V^{ }_{j}} \hat L_{V^{\dagger}_{j}}-
\hat R_{V^{\dagger}_{j}} \hat R_{V^{ }_{j}} \Bigr). \]
In the case of a n-level system ($dim{\cal H}=n$), evolution
equation (\ref{L2}) can be given in the form \cite{GKS}:
\begin{equation} \label{GKSe}
\frac{d}{dt} \rho(t)=-i[H,\rho(t)]+\sum^{n^2-1}_{k,l=1}
C_{kl} \Bigl( F_k \rho(t) F^{\dagger}_l-
\rho(t) \circ (F_k F^{\dagger}_l) \Bigr) \ , \end{equation}
where
\[ H^{\dagger}=H, \quad Tr(H)=0, \quad Tr(F_k)=0 \ , \quad
Tr(F^{\dagger}_k F_l)=\delta_{kl}, \]
The matrix $\{C_{kl}\}$ is a positive matrix $(n^2-1)\times (n^2-1)$
and $\{I,F_k|k=1,...,n^2-1\}$ is an operator basis for the space
of bounded operators on ${\cal H}_{n}$.
The matrix $\{C_{kl}\}$ is called a positive matrix if all
elements $C_{kl}$ are real ($C^{*}_{kl}=C_{kl}$) and positive $C_{kl}>0 $.
For the proofs of (\ref{GKSe}), we refer to \cite{GKS}.

For a given $\hat \Lambda$, operator $H$ is uniquely determined by the
condition $Tr(H)=0$, and the matrix $\{C_{kl}\}$ is uniquely determined
by the choice of the $F_k$. The conditions $Tr(H)=0$ and $Tr(F_k)=0$
provide a canonical separation of the superoperator $\hat \Lambda$ into
a Hamiltonian plus dissipative part.

If the condition of completely positivity is replaced by the
weaker requirement of simple positivity, the generator for a
n-level system can be again be written in the form (\ref{GKSe}),
where the matrix $\{C_{kl}\}$ is a matrix of positive defined
Hermitian form \cite{BZPP,GFVKS}, i.e.
\[ \sum^{n^{2}-1}_{kl=1} C_{kl} z_k z^*_l > 0, \]
for all $z_k \in \mathbb{C}$.
The matrix $\{C_{kl}\}$ of Hermitian form is Hermitian matrix
($C^{*}_{kl}=C_{lk}$).
It is known \cite{Gant} that Hermitian form is positive if and only if
$$det\left(
\begin{array}{cccc}
C_{11}&C_{12}&...&C_{1k}\\
C_{21}&C_{22}&...&C_{2k}\\
.&...&...&...\\
C_{k1}&C_{k2}&...&C_{kk}\\
\end{array}
\right)>0,$$
for all $ k=1,2,...,n^{2}-1$.
This condition is equivalent to condition of positivity
for matrix eigenvalues.

Let us consider a two-level quantum system (qubit) \cite{GKS,,GFVKS,PZ,Kos}
for a positive trace-preserving semigroup.
Let $\{F_{\mu}\}$, where $\mu \in \{0,1,2,3\}$,
be a complete orhonormal  set of self-adjoint matrices:
\[ F_0=\frac{1}{\sqrt{2}}I, \quad F_1=\frac{1}{2}\sigma_1, \quad
F_2=\frac{1}{\sqrt{2}}\sigma_2, \quad F_3=\frac{1}{\sqrt{2}}\sigma_3. \]
Let Hamilton operator $H$ and state $\rho(t)$ have the form
\[ H=\sum^{3}_{k=1} H_k \sigma_k\ , \quad
\rho(t)=\frac{1}{2}(P_0 I+P_k(t) \sigma_k), \]
where $P_0=1$ in all cases. Using the relations
\[ \sigma_k \sigma_l=I\delta_{kl}+i\sum^{3}_{m=1}
\varepsilon_{klm} \sigma_m, \quad
[\sigma_k, \sigma_l]=2i\sum^{3}_{m=1}
\varepsilon_{klm} \sigma_m, \]
and $\varepsilon_{klm}\varepsilon_{ijm}=
\delta_{ki}\delta_{lj}-\delta_{kj}\delta_{li}$,
for (\ref{GKSe}) we obtain the equations:
\[ \frac{d}{dt}P_{k}(t)=
\sum^{3}_{l=1}\Bigl( 2H_m\varepsilon_{kml}+
\frac{1}{8}(C_{kl}+C_{lk})-\frac{1}{4}C\delta_{kl} \Bigr) P_l(t)-\]
\[-\frac{1}{4}\varepsilon_{ijk}(ImC_{ij}) P_0, \]
where $C=\sum^{3}_{m=1}C_{mm}$ and $k,l \in \{1,2,3\}$. We can
rewrite this equation in the form
\begin{equation} \label{P} \frac{d}{dt}P_{\mu}(t)=\sum^{3}_{\nu=0}
{\cal L}_{\mu \nu} P_{\nu}(t) \ ,  \end{equation}
where $\mu,\nu \in \{0,1,2,3\}$ and the matrix ${\cal L}_{\mu \nu}$ is
{\small
$$
\left(
\begin{array}{cccc}
0&0&0&0\\
B_1&-C_{(22)}-C_{(33)}&C_{(12)}-2H_{3}&C_{(13)}+2H_{2}\\
B_2&C_{(12)}+2H_{3}&-C_{(11)}-C_{(33)}&C_{(23)}-2H_{1}\\
B_3&C_{(13)}-2H_{2}&C_{(23)}+2H_{1}&-C_{(11)}-C_{(22)}
\end{array}
\right),$$
}
where
\[ B_k=-\frac{1}{4}\varepsilon_{ijk}(ImC_{ij}) \ ,
\quad C_{(kl)}=\frac{1}{8}(C_{kl}+C_{lk}). \] If the matrix
$C_{kl}$ and Hamilton operator $H$ are not time-dependent, then
equation (\ref{P}) has a solution
\[ P_{\mu}(t)=\sum^{3}_{\nu=0}{\cal E}_{\mu \nu}(t,t_0) P_\nu(t_0), \]
where the matrix ${\cal E}_{\mu \nu}$ is
$${\cal E}=\left(
\begin{array}{cccc}
1&0&0&0\\
T_1&R_{11}&R_{12}&R_{13}\\
T_2&R_{21}&R_{22}&R_{23}\\
T_3&R_{31}&R_{32}&R_{33}
\end{array}
\right).$$
The matrices $T$ and $R$ of the matrix ${\cal E}_{\mu \nu}$ are defined by
\[ T=(e^{\tau A}-I) (\tau A)^{-1}B=\sum^{\infty}_{n=0}
\frac{\tau^{n-1}}{n!} A^{n-1}, \]
\[ R=e^{\tau A}=\sum^{\infty}_{n=0} \frac{\tau^{n}}{n!} A^{n}, \]
where $\tau=t-t_{0}$ and elements of the matrix $A$ are
\[ A_{kl}=2H_m\varepsilon_{kml}+
\frac{1}{8}(C_{kl}+C_{lk})-\frac{1}{4}C\delta_{kl}. \]
If $C_{kl}$ is a real matrix, then all $T_k=0$, where $k=1,2,3$.

\subsection{Quantum operation}

Unitary evolution (\ref{UU}) is not the most general type
of state change possible for quantum systems.
A most general state change of a quantum system
is a positive map ${\cal E}$ which
is called a quantum operation or superoperator.
For the concept of quantum operations see \cite{Kr1,Kr2,Kr3,Kr4,Schu}.
In the formalism of quantum operations the final (output)
state $\rho^{\prime}$ is related to the initial (output) state $\rho$
by a map
\begin{equation} \label{Et} \rho \ \rightarrow \ \rho^{\prime}=
\frac{{\cal E}(\rho)}{Tr({\cal E}(\rho) )} \ . \end{equation} The
trace in the denominator is induced in order to preserve the trace
condition, $Tr(\rho^{\prime})=1$. In general case, this
denominator leads to the map is nonlinear, where the map ${\cal E}$
is a linear positive map.

The quantum operation ${\cal E}$
usually considered as a completely positive map \cite{AL}.
The most general form for completely positive quantum
operation ${\cal E}$ is
\[ {\cal E}(\rho)=\sum^{m}_{j=1} A_j \rho A^{\dagger}_j. \]
By definition, $Tr({\cal E}(\rho))$ is the probability that the process
represented by ${\cal E}$ occurs, when $\rho$ is the initial state.
The probability never exceed 1. The quantum operation ${\cal E}$
is trace-decreasing, i.e. $Tr({\cal E}(\rho)) \le 1$ for all
density matrix operators $\rho$. This condition can be expressed
as an operator inequality for $A_{j}$.
The operators $A_j$ must satisfy
\[ \sum^{m}_{j=1} A^{\dagger}_j A_{j} \le I. \]
The normalized post-dynamics system state is defined by (\ref{Et}).
The map (\ref{Et}) is nonlinear trace-preserving map.
If the linear quantum operation ${\cal E}$ is trace-preserving
$Tr({\cal E}(\rho))=1$, then
\[ \sum^{m}_{j=1} A^{\dagger}_j A_{j}=I. \]
Notice that a trace-preserving quantum operation
${\cal E}(\rho)=A\rho A^{\dagger}$ must be a unitary
transformation ($A^{\dagger}A=AA^{\dagger}=I$).

The example of nonunitary dynamics is associated with the
measurement of quantum system. The system being measured is no
longer a closed system, since it is interacting with the measuring
device. The usual way \cite{Dav} to describe a measurement (von
Neumann measurement) is a set of projectors $P_{k}$ onto the pure
state space of the system such that
\[ P_{k}P_{l}=\delta_{kl}P_{k} \ , \quad P^{\dagger}_{k}=P_{k} \ ,
\quad \sum_{k}P_{k}=I. \]
The unnormalized state of the system after the measurement is
given by
\[ {\cal E}_{k}(\rho)=P_{k}\rho P_{k}. \]
The probability of this measurement result is given by
\[ p(k)=Tr( {\cal E}_{k}(\rho) ). \]
The normalization condition, $\sum_{k}p(k)=1$ for all density
matrix operators, is equivalent to the completeness condition
$\sum_{k}P_{k}=I$.
If the state of the system before the measurement was $\rho$,
than the normalized state of the system after the measurement is
\[ \rho'=p^{-1}(k){\cal E}_{k}(\rho). \]


\section{Quantum four-valued logic gates}


In this section we consider some properties of four-valued
logic gates. We connect quantum
four-valued logic gates with unitary two-valued logic gates
and quantum operations by the generalized computational basis.


\subsection{Generalized quantum gates}

Quantum operations can be considered as generalized quantum gates
act on general (mixed) states.
Let us define a quantum 4-valued logic gates.

\noindent {\bf Definition}
{\it Quantum four-valued logic gate is
a superoperator $\hat{\cal E}$ on Liouville space
$\overline{\cal H}^{(n)}$ which maps a density
matrix operator $|\rho)$ of $n$-ququats to a density matrix operator
$|\rho')$ of $n$-ququats.}

A generalized quantum gate is a superoperator $\hat{\cal E}$ which
maps density matrix operator $|\rho)$ to density matrix operator
$|\rho^{\prime})$.
If $\rho$ is operator of density matrix, then $\hat{\cal E}(\rho)$
should also be a density matrix operator.
Any density matrix operator is self-adjoint ($\rho^{\dagger}(t)=\rho(t)$),
positive ($\rho(t)>0$) operator with unit trace ($ Tr\rho(t)=1$).
Therefore we have some requirements for superoperator $\hat{\cal E}$.

The requirements for a superoperator $\hat{\cal E}$
to be a generalized quantum gate are as follows: \\
1. The superoperator $\hat{\cal E}$ is {\it real} superoperator, i.e.
$\Bigl(\hat{\cal E}(A)\Bigr)^{\dagger}=\hat{\cal E}(A^{\dagger})$
for all $A$ or $\Bigl(\hat{\cal E}(\rho)\Bigr)^{\dagger}=\hat{\cal E}(\rho)$.
Real superoperator $\hat{\cal E}$ maps self-adjoint operator
$\rho$ into self-adjoint operator $\hat{\cal E}(\rho)$:
$ (\hat{\cal E}(\rho))^{\dagger}=\hat{\cal E}(\rho)$.\\
2.1. The gate $\hat{\cal E}$ is a {\it positive} superoperator,
i.e. $\hat{\cal E}$ maps positive operators to positive operators:
\ $\hat{\cal E}(A^{2}) >0$ for all $A\not=0$ or $\hat{\cal E}(\rho)\ge 0$.\\
2.2. We have to assume the superoperator $\hat{\cal E}$
to be not merely positive but completely positive. The
superoperator $\hat{\cal E}$ is {\it completely positive} map of
Liouville space, i.e. the positivity is remained if we extend the
Liouville space $\overline{\cal H}^{(n)}$ by adding more qubits.
That is, the superoperator $\hat{\cal E} \otimes \hat I^{(m)}$
must be positive, where $\hat I^{(m)}$ is the identity
superoperator on some Liouville space $\overline{\cal H}^{(m)}$.\\
3. The superoperator $\hat{\cal E}$ is {\it trace-preserving} map, i.e.
\[ (I|\hat{\cal E}|\rho)=(\hat{\cal E}^{\dagger}(I)|\rho)=1 \quad
or \quad \hat{\cal E}^{\dagger}(I)=I. \]
3.1. The superoperator $\hat{\cal E}$ is a {\it convex linear} map
on the set of density matrix operators, i.e.
\[ \hat{\cal E}(\sum_{s} \lambda_{s} \rho_{s})=
\sum_{s} \lambda_{s} \hat{\cal E}(\rho_{s}), \]
where all $\lambda_{s}$ are $0<\lambda_{s}<1$ and
$\sum_{s} \lambda_{s}=1$.
Note that any convex linear map of density matrix operators
can be uniquely extended to a {\it linear} map on Hermitian operators.
Any linear completely positive superoperator can be represented by
\[ \hat{\cal E}=\sum^{m}_{j} \hat L_{A_{j}} \hat R_{A^{\dagger}_{j}}. \]
If $\hat{\cal E}$ is trace-preserving superoperator, then
\[ (I|\hat{\cal E}|\rho)=1 \quad or \quad
\hat{\cal E}^{\dagger}(I)=I. \]
3.2. The restriction to linear gates is unnecessary.
Let us consider $\hat{\cal E}$ is a linear superoperator which is not
trace-preserving. This superoperator is not a quantum gate.
Let $(I|\hat{\cal E}|\rho)=Tr(\hat{\cal E}(\rho))$
is a probability that the process represented
by the superoperator $\hat{\cal E}$ occurs.
Since the probability is nonnegative and
never exceed 1, it follows that the superoperator
$\hat{\cal E}$ is a trace-decreasing superoperator:
\[ 0\le (I|\hat{\cal E}|\rho) \le 1 \quad
or \quad \hat{\cal E}^{\dagger}(I) \le I. \]
In general case, the linear trace-decreasing superoperator is not a
quantum four-valued logic gate, since it can be not trace-preserving.
The generalized quantum gate can be defined as {\it nonlinear
trace-preserving} gate $\hat{\cal N}$ by
\[ \hat{\cal N}|\rho)=
( \hat{\cal E}|\rho)I|\hat{\cal E}|\rho)^{-1} \quad or
\quad \hat{\cal N}(\rho)=
\frac{\hat{\cal E}(\rho)}{Tr(\hat{\cal E}(\rho))}, \]
where $\hat{\cal E}$ is a linear completely positive trace-decreasing
superoperator.

In the generalized computational basis the gate
$\hat{\cal E}$ can be represented by
\[ \hat{\cal E}=\frac{1}{2^n}\sum^{N-1}_{\mu=0}
\sum^{N-1}_{\nu=0}
{\cal E}_{\mu \nu} |\sigma_{\mu})(\sigma_{\nu}|. \]
where $N=4^n$, $\mu$ and $\nu$ are 4-valued representation of
\[ \mu=\mu_1 4^{N-1}+...+\mu_{N-1}4+\mu_N , \]
\[ \nu=\nu_1 4^{N-1}+...+\nu_{N-1}4+\nu_N, \]
\[ \sigma_{\mu}=\sigma_{\mu_1} \otimes...\otimes \sigma_{\mu_m}, \]
$\mu_i,\nu_i \in \{0,1,2,3\}$ and
${\cal E}_{\mu \nu}$ are elements of some matrix.

\subsection{General quantum operation as four-valued logic gates}

\bp
{\it In the generalized computational basis $|\mu)$ any linear
two-valued logic quantum operation ${\cal E}$ can be represented as
a  quantum four-valued logic gate $\hat{\cal E}$ defined by
\[ \hat{\cal E}=\sum^{N-1}_{\mu=0}\sum^{N-1}_{\nu=0}
{\cal E}_{\mu \nu} \ |\mu)(\nu|  \ , \]
where
\[ {\cal E}_{\mu \nu}=\frac{1}{2^n}
Tr\Bigl(\sigma_{\mu} \hat{\cal E} (\sigma_{\nu}) \Bigr), \]
and  $\sigma_{\mu} =\sigma_{\mu_1} \otimes ...  \otimes \sigma_{\mu_n}$.}
\ep

\noindent {\bf Proof.}
The state $\rho(t)$ in the generalized computational basis
$|\mu)$ has the form
\[ |\rho(t))=\sum^{N-1}_{\mu=0} |\mu)\rho_{\mu}(t)  \ , \]
where $N=4^{n}$ and
\[ \rho_{\mu}(t)=(\mu|\rho(t))=
\frac{1}{\sqrt{2^n}}Tr(\sigma_{\mu}\rho(t)). \]
The quantum operation ${\cal E}$ define a four-valued logic gate by
\[ |\rho(t))= \hat{\cal E}_{t}|\rho) =| {\cal E}_{t}(\rho) )=
\sum^{N-1}_{\nu=0} |{\cal E}_{t}(\sigma_{\nu}) )
\frac{1}{\sqrt{2^n}} \rho_{\nu}(t_{0}). \]
Then
\[ (\mu|\rho(t))=\sum^{N-1}_{\nu =0}
(\sigma_{\mu}|{\cal E}_{t}( \sigma_{\nu}) )
\frac{1}{2^n} \rho_{\nu}(t_{0}). \]
Finally, we obtain
\[ \rho_{\mu}(t)=\sum^{N-1}_{\nu=0}
{\cal E}_{\mu \nu}\rho_{\nu}(t_0), \]
where
\[ {\cal E}_{\mu \nu}=\frac{1}{2^n}
(\sigma_{\mu}|{\cal E}_{t}(\sigma_{\nu}) )
=\frac{1}{2^n}Tr\Bigl(\sigma_{\mu} {\cal E}_{t}(\sigma_{\nu}) \Bigr). \]
This formula defines a relation between general quantum
operation ${\cal E}$ and the real $4^{n}\times 4^{n}$ matrix
${\cal E}_{\mu \nu}$ of  four-valued logic gate $\hat{\cal E} $.


Four-valued logic gates $\hat{\cal E}$
in the matrix representation are represented by $4^{n}\times 4^{n}$
matrices ${\cal E}_{\mu \nu}$.
The matrix ${\cal E}_{\mu \nu}$ of the gate $\hat{\cal E}$ is
$${\cal E}=\left(
\begin{array}{cccc}
{\cal E}_{00}&{\cal E}_{01}&...&{\cal E}_{0a}\\
{\cal E}_{10}&{\cal E}_{11}&...&{\cal E}_{1a}\\
...&...& ...&...\\
{\cal E}_{a0}&{\cal E}_{a1}&...
&{\cal E}_{aa}
\end{array}
\right),$$
where $a=N-1=4^{n}-1$.

In matrix representation the gate $\hat{\cal E}$ maps
the state $|\rho)=\sum^{N-1}_{\nu=0}|\nu) \rho_{\nu}$
to the state $|\rho^\prime)=\sum^{N-1}_{\mu}|\mu) \rho^{\prime}_{\mu}$ by
\begin{equation} \label{rEr} \rho^{\prime}_{\mu}=\sum^{N-1}_{\nu=0}
{\cal E}_{\mu \nu} \rho_{\nu} \ . \end{equation}
where $\rho^{\prime}_{0}=\rho_{0}=1/\sqrt{2^{n}}$.
It can be written in the form
$$\left(
\begin{array}{c}
\rho^{\prime}_0\\
\rho^{\prime}_1\\
...\\
\rho^{\prime}_{a}
\end{array}
\right)=
\left(
\begin{array}{ccccc}
{\cal E}_{00}&{\cal E}_{01}&...&{\cal E}_{0a}\\
{\cal E}_{10}&{\cal E}_{11}&...&{\cal E}_{1a}\\
...&...& ...&...\\
{\cal E}_{a0}&{\cal E}_{a1}&...
&{\cal E}_{aa}
\end{array}
\right)
\left(
\begin{array}{c}
\rho_0\\
\rho_1\\
...\\
\rho_{a}
\end{array}
\right).$$

Since $P_{\mu}=\sqrt{2^{n}} \rho_{\mu}$ and
$P^{\prime} _{\mu}=\sqrt{2^{n}} \rho^{\prime}_{\mu}$,
it follows that representation (\ref{rEr}) for linear gate
$\hat{\cal E}$ is equivalent to
\begin{equation} \label{PEP} P^{\prime}_{\mu}=\sum^{N-1}_{\nu=0}
{\cal E}_{\mu \nu} P_{\nu} \ . \end{equation}
It can be written in the form
$$\left[
\begin{array}{c}
P^{\prime}_0\\
P{\prime}_1\\
...\\
P^{\prime}_{a}
\end{array}
\right]=
\left(
\begin{array}{cccc}
{\cal E}_{00}&{\cal E}_{01}&...&{\cal E}_{0a}\\
{\cal E}_{10}&{\cal E}_{11}&...&{\cal E}_{1a}\\
...&...& ...&...\\
{\cal E}_{a0}&{\cal E}_{a1}&...
&{\cal E}_{aa}
\end{array}
\right)
\left[
\begin{array}{c}
P_0\\
P_1\\
...\\
P_{a}
\end{array}
\right].$$
where $P_{0}=1$.
Note that if we use different matrix representation of state
$|\rho)$ or $|\rho]$ we can use identical matrices
representation for gate $\hat{\cal E}$.

\bp
{\it In the generalized computational basis $|\mu)$ the matrix
${\cal E}_{\mu \nu}$ of general quantum four-valued
logic gate
\begin{equation}  \label{elr}
\hat{\cal E}=\sum^{m}_{j=1} \hat L_{A_j} \hat R_{A^{\dagger}_{j}}
\end{equation}
is real
${\cal E}^{*}_{\mu \nu}={\cal E}_{\mu \nu}$. }
\ep

\noindent {\bf Proof.}
\[ {\cal E}_{\mu \nu}=\frac{1}{2^n}\sum^{m}_{j=1}
Tr\Bigl(\sigma_{\mu} A_{j} \sigma_{\nu} A^{\dagger}_{j} \Bigr)=
\frac{1}{2^n} \sum^{m}_{j=1}
(A^{\dagger}_j  \sigma_{\mu}| \sigma_{\nu}A^{\dagger}_j ). \]
\[ {\cal E}^{*}_{\mu \nu}=\frac{1}{2^n} \sum^{m}_{j=1}
(A^{\dagger}_j \sigma_{\mu}| \sigma_{\nu} A^{\dagger}_j)^{*}=
\frac{1}{2^n} \sum^{m}_{j=1}
( \sigma_{\nu} A^{\dagger}_j|A^{\dagger}_j  \sigma_{\mu} )=  \]
\[ =\frac{1}{2^n} \sum^{m}_{j=1}
Tr\Bigl(A_j \sigma_{\nu} A^{\dagger}_{j} \sigma_{\mu} \Bigr)=
\frac{1}{2^n} \sum^{m}_{j=1}
Tr\Bigl( \sigma_{\mu} A_{j} \sigma_{\nu} A^{\dagger}_j \Bigr)=
{\cal E}_{\mu \nu}. \]

\bp
{\it Any real matrix ${\cal E}_{\mu \nu}$ associated with
linear (trace-preserving) quantum four-valued
logic gates (\ref{elr}) has}
\[ {\cal E}_{0 \nu}=\delta_{0 \nu}. \]
\ep

\noindent {\bf Proof.}
\[ {\cal E}_{0 \nu}=\frac{1}{2^{n}}
Tr\Bigl(\sigma_{0}{\cal E}(\sigma_{\nu}) \Bigr)=
\frac{1}{2^{n}} Tr\Bigl({\cal E}(\sigma_{\nu}) \Bigr)= \]
\[ =\frac{1}{2^{n}}
Tr\Bigl(\sum^{m}_{j=1} A_j \sigma_{\nu}A^{\dagger}_j \Bigr)=
\frac{1}{2^{n}} Tr\Bigl((\sum^{m}_{j=1}
A^{\dagger}_j A_j )\sigma_{\nu} \Bigr)=\]
\[=\frac{1}{2^{n}} Tr \sigma_{\nu}=\delta_{0\nu}. \]
The general linear n-ququats quantum gate has the form:
$${\cal E}=\left(
\begin{array}{ccccc}
1&0&0&...&0\\
T_{1}&R_{11}&R_{12}&...&R_{1 \ N-1 }\\
T_{2}&R_{21}&R_{22}&...&R_{2 \ N-1}\\
...&...&...& ...&...\\
T_{N-1}&R_{N-1 \ 1}&R_{N-1 \ 2}&...&R_{N-1 \ N-1}
\end{array}
\right).$$
Completely positive condition leads to some inequalities
\cite{KR,RSW,Choi} for matrix elements ${\cal E}_{\mu \nu}$.

In general case, linear quantum 4-value logic gate acts on
$|0)$ by
\[ \hat{\cal E}|0)=|0)+\sum^{N-1}_{k=1}T_{k}|k). \]
For example, single ququat quantum gate acts by
\[ \hat{\cal E}|0)=|0)+T_{1}|1)+T_{2}|2)+T_{3}|3). \]
If all $T_{k}, \ k=1,...,N-1$ is equal to zero,
then $\hat{\cal E}|0)=|0)$. The linear quantum gates with $T=0$
conserve the maximally mixed state $|0]$ invariant.

\noindent {\bf Definition} {\it A quantum four-valued logic gate
$\hat{\cal E}$ is called unital gate or gate with $T=0$ if
maximally mixed state $|0]$ is invariant under the action of this
gate: $\hat{\cal E}|0]=|0]$.}

The output state of a linear quantum four-valued logic gate
$\hat{\cal E}$ is $|00...0]$
if and only if the input state is $|00...0]$.
If $\hat{\cal E}|00...0]\not=|00...0]$, then
$\hat{\cal E}$ is not unital gate.

\bp
{\it The matrix ${\cal E}_{\mu \nu}$ of linear trace-preserving
n-ququats gate $\hat{\cal E}$ is an element of group $TGL(4^n-1,\mathbb{R})$
which is a semidirect product of general linear group
$GL(4^n-1,\mathbb{R})$ and translation group $T(4^n-1,\mathbb{R})$.}
\ep

\noindent {\bf Proof.}
This proposition follows from proposition 3.
Any element (gate matrix ${\cal E}_{\mu \nu}$) of group $TGL(4^n-1,\mathbb{R})$
can be represented by
$${\cal E}(T,R)=\left(
\begin{array}{cc}
1&0\\
T&R
\end{array}
\right),$$
where $T$ is a column with $4^n-1$ elements,
$0$ is a line with $4^n-1$ zero elements, and
$R$ is a real $(4^n-1)\times (4^n-1)$ matrix
$R \in GL(4^n-1,\mathbb{R})$.
If $R$ is orthogonal $(4^{n}-1)\times(4^{n}-1)$ matrix ($R^TR=I$),
then we have motion group \cite{Vil,Vil1,Or}.
The group multiplication of elements ${\cal E}(T,R)$ and ${\cal E}(T',R')$
is defined by
\[ {\cal E}(T,R){\cal E}(T',R')={\cal E}(T+RT',RR'). \]
In particular, we have
\[ {\cal E}(T,R)={\cal E}(T,I){\cal E}(0,R) \ , \quad
{\cal E}(T,R)={\cal E}(0,R){\cal E}(R^{-1}T,I). \]
where $I$ is unit $(4^n-1)\times (4^n-1)$ matrix.

Therefore any linear quantum gate can be decompose on
unital gate and translation gate.
It allows to consider two types of linear trace-preserving gates:\\
1) Translation gates $\hat{\cal E}^{(T)}$
defined by matrices ${\cal E}(T,I)$.\\
2) Unital quantum gates $\hat{\cal E}^{(T=0)}$ with
the matrices ${\cal E}(0,R)$.

Translation gate $\hat{\cal E}^{(T)}$ is
\[ \hat{\cal E}^{(T)}=\sum^{N-1}_{\mu=0} |\mu)(\mu|+
\sum^{N-1}_{k=1}T_k |k)(0|, \]
has the matrix
$${\cal E}(T,I)=\left(
\begin{array}{ccccc}
1&0&0&...&0\\
T_{1}&1&0&...&0\\
T_{2}&0&1&...&0\\
...&...&...& ...&...\\
T_{N-1}&0&0&...&1
\end{array}
\right).$$
One-parameter subgroups $T(4^n-1,\mathbb{R})$ of n-ququats translation
gates consist of one-parameters $4^n-1$ gates
\[ \hat{\cal E}^{(T,k)}(t)=\sum^{N-1}_{\mu=0} |\mu)(\mu|+
t |k)(0|, \] where $t$ is a real parameter and
$k=1,2,...,4^{n}-1$. Generators of the gates are defined by
\[ \hat H_k=\Bigl(\frac{d}{dt} \hat{\cal E}^{(T,k)}(t) \Bigr)_{t=0}
=|k)(0|. \]

The quantum n-ququats unital gate can be represented by
\[ \hat{\cal E}^{(T=0)}=|0)(0|+\sum^{N-1}_{k=1}
\sum^{N-1}_{l=1} R_{kl}|k)(l|, \]
where $N=4^{n}$.
The gate matrix ${\cal E}(0,R)$ has the form
$${\cal E}(0,R)=\left(
\begin{array}{ccccc}
1&0&0&...&0\\
0&R_{11}&R_{12}&...&R_{1 \ N-1 }\\
0&R_{21}&R_{22}&...&R_{2 \ N-1}\\
...&...&...& ...&...\\
0&R_{N-1 \ 1}&R_{N-1 \ 2}&...&R_{N-1 \ N-1}
\end{array}
\right).$$
In matrix representation the linear trace-preserving gates with $T=0$
can be described by group $GL(4^n-1,\mathbb{R})$
which define a set of all
linear transformations of $\overline{\cal H}^{(n)}$ by
(\ref{rEr}) or (\ref{PEP}).
The group $GL(4^n-1,\mathbb{R})$ has $(4^{n}-1)^2$
independent one-parameter subgroups $GL_{kl}(4^n-1,\mathbb{R})$
of one-parameter gates $\hat{\cal E}_{(kl)}(t)$ such that
\[ \hat{\cal E}^{(kl)}(t)=|0)(0|+t|k)(l|. \]
Generators are defined by
\[ \hat H_{kl}=\Bigl(\frac{d}{dt} \hat{\cal E}^{(kl)}(t) \Bigr)_{t=0}
=|k)(l|. \]
where $k,l=1,2,...,4^{n}-1$.
The generators $\hat H_{kl}$ of the one-parameter subgroup
$GL_{kl}(4^n-1,\mathbb{R})$ are represented by
$4^n \times 4^n$ matrix $H_{kl}$ with elements
\[ (H_{kl})_{\mu \nu}=\delta_{\mu k} \delta_{\nu l}. \]
The set of superoperators $\{\hat H_{kl}\}$ is a basis
of Lie algebra $gl(4^n-1,\mathbb{R})$ such that
\[ [\hat H_{ij},\hat H_{kl}]=
\delta_{jk} \hat H_{il}-\delta_{il} \hat H_{jk}. \]


\subsection{Decomposition for linear quantum gates}

Let us consider the n-ququats linear gate
\begin{equation} \label{LGE}
\hat{\cal E}=|0)(0|+\sum^{N-1}_{\mu=1} T_{\mu} |\mu)(0|+
\sum^{N-1}_{\mu=1} \sum^{N-1}_{\nu=1} R_{\mu \nu} |\mu)(\nu|,
\end{equation}
where $N=4^{n}$. The gate matrix ${\cal E}(T,R)$ is an element of Lie
group $TGL(4^{n}-1,\mathbb{R})$. The  matrix $R$
is an element of Lie group $GL(4^{n}-1,\mathbb{R})$.

\noindent {\bf Theorem 1.} (Singular Valued Decomposition for Matrix)\\
{\it Any real  matrix $R$ can be written in the form
$R={\cal U}_{1} D {\cal U}^{\small T}_{2},$
where
${\cal U}_{1}$ and ${\cal U}_{2}$  are
real orthogonal $(N-1)\times(N-1)$ matrices and
$D=diag(\lambda_1,...,\lambda_{N-1})$
is diagonal $(N-1)\times(N-1)$  matrix
such that
\quad
$\lambda_1 \ge \lambda_2 \ge ... \ge \lambda_{N-1} \ge 0$.

}

\noindent {\bf Proof.}
This theorem is proved  in \cite{EY,Lan,Sc,Gant}.

Let us consider the unital gates $\hat{\cal E}^{(T=0)}$
defined by (\ref{LGE}), where all $T_\mu=0$.

\noindent {\bf Theorem 2.} (Singular Valued Decomposition for Gates)\\
{\it Any unital linear gate $\hat{\cal E}$ defined by (\ref{LGE})
with all $T_{\mu}=0$ can be represented by
\[ \hat{\cal E}=\hat{\cal U}_{1} \ \hat D \ \hat{\cal U}_{2}, \]
where\\
$\hat{\cal U}_{1}$ and $\hat{\cal U}_{2}$ are
unital orthogonal quantum gates
\begin{equation} \label{Ui}
\hat{\cal U}_{i}=|0)(0|+\sum^{N-1}_{\mu=1} \sum^{N-1}_{\nu=1}
{\cal U}^{(i)}_{\mu \nu} |\mu)(\nu|,
\end{equation}
$\hat D$ is a unital diagonal quantum gate, such that
\begin{equation} \label {D} \hat D=|0)(0|+\sum^{N-1}_{\mu=1}
\lambda_{\mu} |\mu)(\mu|, \end{equation}
where $\lambda_{\mu} \ge 0$.
}

\noindent {\bf Proof.} The proof of this theorem can be easy realized
in matrix representation by using theorem 1.

In general case, we have the following theorem.

\noindent {\bf Theorem 3.} (Singular Valued Decomposition for Gates)\\
{\it Any linear quantum four-valued logic gate (\ref{LGE})
can be represented by
\[ \hat{\cal E}=\hat{\cal E}^{(T)} \hat{\cal U}_{1} \
\hat D \ \hat{\cal U}_{2}, \]
where\\
$\hat{\cal U}_{1}$ and $\hat{\cal U}_{2}$ are
unital orthogonal quantum gates (\ref{Ui}).\\
$\hat D$ is a unital diagonal quantum gate (\ref{D}).\\
$\hat{\cal E}^{(T)}$ is a translation quantum gate, such that
\[ \hat{\cal E}^{(T)}=|0)(0|+\sum^{N-1}_{\mu=1}
|\mu)(\mu|+\sum^{N-1}_{\mu=1} T_{\mu} |\mu)(0|. \]
}

\noindent {\bf Proof.} The proof of this theorem can be easy realized
in matrix representation by using Proposition 4 and Theorem 1.


As a result we have that any trace-preserving gate can be realized
by 3 types of gates: (1) unital orthogonal quantum gates
$\hat{\cal U}$ with matrix ${\cal U}\in SO(4^{n}-1,\mathbb{R})$; (2)
unital diagonal quantum gate $\hat D$ with matrix $D \in
D(4^{n}-1,\mathbb{R})$; (3) nonunital translation gate $\hat{\cal
E}^{(T)}$ with matrix ${\cal E}^{(T)}\in T(4^{n}-1,\mathbb{R})$.



\bp
{\it If the quantum operation ${\cal E}$ has the form
\[ {\cal E}(\rho)=\sum^{m}_{j=1} A_{j} \rho A^{\dagger}_{j}, \]
where $A$ is a self-adjoint operator ($A^{\dagger}_{j}=A_{j}$), then
quantum four-valued logic gate $\hat{\cal E}$ is described
by symmetric matrix ${\cal E}_{\mu \nu}={\cal E}_{\nu \mu}$.}
\ep

\noindent {\bf Proof.} If $A^{\dagger}_j=A_j$, then
\[ {\cal E}_{\mu \nu}=
\frac{1}{\sqrt{2^{n}}}\sum^{m}_{j=1}
Tr (\sigma_{\mu}A_{j} \sigma_{\nu} A_{j})=\]
\[=\frac{1}{\sqrt{2^{n}}} \sum^{m}_{j=1}
Tr (\sigma_{\nu}A_{j} \sigma_{\mu} A_{j})={\cal E}_{\nu \mu}. \]
This gate is trace-preserving if
${\cal E}_{\mu 0}={\cal E}_{0 \mu}=\delta_{\mu 0}$.

The symmetric n-ququat linear (trace-preserving)
quantum gate has the form
\begin{equation} \label{ES} \hat{\cal E}^{(S)}=|0)(0|+
\sum^{N-1}_{\mu=1} \sum^{N-1}_{\nu=1} S_{\mu \nu}|\mu)(\nu|. 
\end{equation}
where $S_{\mu \nu}=S_{\nu \mu}$ and this gate is unital 
($T_k=0$ for all $k$).

\noindent {\bf Theorem 4.} (Polar Decomposition for matrix)\\
{\it Any real $(N-1)\times(N-1)$ matrix $R$ can be written
in the form $R={\cal U} S$ or
$R=S^{\prime} {\cal U}^{\prime}$, where
${\cal U}$ and ${\cal U}^{\prime}$ are
orthogonal $(N-1)\times(N-1)$ matrices and
$S$ and $S^{\prime}$ are symmetric $(N-1)\times(N-1)$
matrices such that
$S=\sqrt{R^{\small T}R}$, $S^{\prime}=\sqrt{RR^{\small T}}$.
}

\noindent {\bf Proof.}
This theorem is proved in \cite{Gant}.

\noindent {\bf Theorem 5.} (Polar Decomposition for gates)\\
{\it Any linear four-valued logic gate (\ref{LGE})
can be written in the form
$\hat{\cal E}=\hat{\cal U} \hat{\cal E}^{(S)}$ or
$\hat{\cal E}=\hat{\cal E}^{(S^{\prime})} \hat{\cal U}^{\prime}$,
where\\
$\hat{\cal U}$ and $\hat{\cal U}^{\prime}$ are orthogonal gates (\ref{Ui}).\\
$\hat{\cal E}^{(S^{\prime})}$ and $\hat{\cal E}^{(S^{\prime})}$
are symmetric gates (\ref{ES}).
}

\noindent {\bf Proof.}
The proof of this theorem can be easy realized
in matrix representation by using Theorem 4.


\subsection{Unitary two-valued logic gates as orthogonal
four-valued logic gates}

Let us rewrite the representation (\ref{rhosigma})
for the mixed state $|\rho(t))$
using generalized computational basis in the form
\[ |\rho(t))=\sum^{N-1}_{\mu=0}|\mu)\rho_{\mu}(t)  \ , \]
where
\[ \rho_{\mu}(t)=(\mu|\rho(t))=\frac{1}{\sqrt{2^n}}
Tr(\sigma_{\mu} \rho(t)). \]
Note that $\rho_0(t)=(0|\rho(t))=1/\sqrt{2^{n}}Tr\rho(t)=1/\sqrt{2^{n}}$
for all cases.

\bp
{\it In the generalized computational basis any unitary two-valued logic
gate $U$ can be considered as a quantum four-valued logic gate:
\begin{equation} \label{P6-1}
\hat{\cal U}=\sum^{N-1}_{\mu=0}\sum^{N-1}_{\nu=0}
{\cal U}_{\mu \nu} |\mu)(\nu|  \ , \end{equation}
where ${\cal U}_{\mu \nu}$ is a real matrix such that
\begin{equation} \label{P6-2}
{\cal U}_{\mu \nu}=\frac{1}{2^n}Tr\Bigl(\sigma_{\nu}U
\sigma_{\mu} U^{\dagger}\Bigr) \ . \end{equation}
}
\ep

\noindent {\bf Proof.}
Let us consider unitary two-valued logic gate $U$.
Using equation (\ref{UU}), we get
\[ |\rho(t))=\hat{\cal U}_{t}|\rho(t_{0}) ). \]
Then
\[ (\mu|\rho(t))=(\mu|\hat{\cal U}_{t}|\rho(t_{0}) )=
\sum^{N-1}_{\nu=0} (\mu|\hat{\cal U}_{t}|\nu)(\nu|\rho(t_{0}) ). \]
Finally, we obtain
\[ \rho_{\mu}(t)=\sum^{N-1}_{\nu=0} {\cal U}_{\mu \nu}(t,t_{0})
\rho_{\nu}(t_{0}), \]
where
\[ {\cal U}_{\mu \nu}(t,t_{0})=(\mu|\hat{\cal U}_{t}|\nu)=
\frac{1}{2^n} (\sigma_{\mu}|{\cal U}_{t}(\sigma_{\nu}) )=\]
\[=\frac{1}{2^n}Tr \Bigl(\sigma_{\mu}U(t,t_{0})
\sigma_{\nu} U^{\dagger}(t,t_{0}) \Bigr). \]
This formula defines a relation between unitary quantum two-valued
logic gates $U$ and the real $4^{n}\times 4^{n}$ matrix ${\cal U}$.

\bp
{\it Any four-valued logic gate associated with unitary 2-valued
logic gate by (\ref{P6-1},\ref{P6-2}) is unital gate, i.e.
gate matrix ${\cal U}$ defined by (\ref{P6-2}) has
${\cal U}_{\mu 0}={\cal U}_{0 \mu }=\delta_{\mu 0}$.}
\ep

\noindent {\bf Proof.}
\[ {\cal U}_{\mu 0}=\frac{1}{2^{n}}
Tr\Bigl(\sigma_{\mu}U\sigma_{0} U^{\dagger}\Bigr)=\frac{1}{2^{n}}
Tr\Bigl(\sigma_{\mu}UU^{\dagger}\Bigr)=\frac{1}{2^{n}}
Tr\sigma_{\nu}. \]
Using $Tr\sigma_{\mu}=\delta_{\mu 0}$
we get ${\cal U}_{\mu 0}=\delta_{\mu 0}$.

Let us denote the gate $\hat{\cal U}$ associated with unitary
two-valued logic gate $U$ by $\hat{\cal E}^{(U)}$.
\bp
{\it If $U$ is unitary two-valued logic gate, then
in the generalized computational basis a quantum four-valued
logic gate $\hat{\cal U}=\hat{\cal E}^{(U)}$ associated with $U$
is represented
by orthogonal matrix ${\cal E}^{(U)}$: }
\begin{equation} \label{ORT}
{\cal E}^{(U)}({\cal E}^{(U)})^{T}=
({\cal E}^{(U)})^{T}{\cal E}^{(U)}=I \ .
\end{equation}
\ep

\noindent {\bf Proof.}
Let $\hat{\cal E}^{(U)}$ is defined by
\[ \hat{\cal E}^{(U)}|\rho)=|U \rho U^{\dagger}) \ ,
\quad \hat{\cal E}^{(U^{\dagger})}|\rho)=|U^{\dagger} \rho U). \]
If $UU^{\dagger}=U^{\dagger}U=I$, then
\[ \hat{\cal E}^{(U)}\hat{\cal E}^{(U^{\dagger})}=
\hat{\cal E}^{(U^{\dagger})}\hat{\cal E}^{(U)}=\hat I. \]
In the matrix representation we have
\[ \sum^{N-1}_{\alpha=0} {\cal E}^{(U)}_{\mu \alpha}
{\cal E}^{(U^{\dagger})}_{\alpha \nu}=
\sum^{N-1}_{\alpha=0} {\cal E}^{(U^{\dagger})}_{\mu \alpha}
{\cal E}^{(U)}_{\alpha \nu}=\delta_{\mu \nu}  \ , \]
i.e. ${\cal E}^{(U^{\dagger})}{\cal E}^{(U)}=
{\cal E}^{(U)}{\cal E}^{(U^{\dagger})}=I$.
Note that
\[ {\cal E}^{(U^{\dagger})}_{\mu \nu}=
\frac{1}{2^{n}}Tr\Bigl( \sigma_{\mu} U^{\dagger}\sigma_{\nu} U \Bigr)=
\frac{1}{2^{n}}Tr\Bigl( \sigma_{\nu} U \sigma_{\mu} U^{\dagger} \Bigr)=
{\cal E}^{(U)}_{\nu \mu}, \]
i.e. ${\cal E}^{(U^{\dagger})}=({\cal E}^{(U)})^{T}$.
Finally, we obtain (\ref{ORT}).

In matrix representation orthogonal gates
can be described by group $SO(4^n-1,\mathbb{R})$
which is a set of all
linear transformations of $\overline{\cal H}^{(n)}$
${\rho'}_{\mu}=\sum^{N-1}_{\nu=0}{\cal E}_{\mu \nu} \rho_{\nu}$ such that
$\sum^{N-1}_{\mu=0} \rho^2_{\mu}=const$ and $det[{\cal E}_{\mu \nu}]=1$.
The group $SO(4^n-1,\mathbb{R})$ has $(4^{n}-1)(2 \ 4^{n-1}-1)$
independent one-parameter subgroups $SO_{kl}(4^n-1,\mathbb{R})$
of one-parameter orthogonal gates which are
\[ \hat{\cal E}^{(kl)}(\alpha)=\sum_{\mu \not=k,l}|\mu)(\mu|+
cos \alpha\Bigl(|k)(k|+|l)(l| \Bigr)+\]
\[+sin \alpha\Bigl(|l)(k|-|k)(l| \Bigr). \]
This gate defines rotation in the flat $(k,l)$.
Let us note that the generators of the one-parameter subgroup
$SO_{kl}(4^n-1,\mathbb{R})$ are represented by antisymmetric
$(4^n-1)\times(4^n-1)$ matrix $X_{kl}$ with elements
\[ (X_{kl})_{\mu \nu}=\delta_{\mu k} \delta_{\nu l}-
\delta_{\mu l} \delta_{\nu k}. \]

\bp
{\it If $\hat{\cal E}^{\dagger}$ is adjoint superoperator for
linear trace-preserving gate $\hat{\cal E}$, then matrices of the
gates are connected by transposition ${\cal E}^{\dagger}={\cal E}^T$:}
\[ ({\cal E}^{\dagger})_{\mu \nu}={\cal E}_{\nu \mu}. \]
\ep

\noindent {\bf Proof.}
Using
\[ \hat{\cal E}=\sum^{m}_{j=1} \hat L_{A_{j}} \hat R_{A^{\dagger}_{j}} \ ,
\quad \hat{\cal E}^{\dagger}=\sum^{m}_{j=1} \hat L_{A^{\dagger}_{j}}
\hat R_{A_{j}}, \]
we get
\[ {\cal E}_{\mu \nu}=\frac{1}{2^{n}} \sum^{m}_{j=1}
Tr (\sigma_{\mu}A_{j}\sigma_{\nu}A^{\dagger}_{j}), \]
\[ ({\cal E}^{\dagger})_{\mu \nu}=\frac{1}{2^{n}} \sum^{m}_{j=1}
Tr (\sigma_{\mu}A^{\dagger}_{j}\sigma_{\nu}A_{j})=\]
\[=\frac{1}{2^{n}} \sum^{m}_{j=1} Tr
(\sigma_{\nu}A_{j}\sigma_{\mu}A^{\dagger}_{j})={\cal E}_{\nu \mu}. \]

Obviously, if
\[ \hat{\cal E}=\sum^{N-1}_{\mu=0}\sum^{N-1}_{\nu=0}
{\cal E}_{\mu \nu}|\mu)(\nu|, \]
then
\[ \hat{\cal E}^{\dagger}=\sum^{N-1}_{\mu=0}
\sum^{N-1}_{\nu=0}{\cal E}_{\nu \mu}|\mu)(\nu|. \]

\bp
{\it If $\hat{\cal E}^{\dagger} \hat{\cal E}=
\hat{\cal E} \hat{\cal E}^{\dagger}= \hat I$,
then $\hat{\cal E}$ is orthogonal gate, i.e.
${\cal E}^T{\cal E}={\cal E}{\cal E}^T=I$.}
\ep

\noindent {\bf Proof.}
If $\hat{\cal E}^{\dagger} \hat{\cal E}=\hat I$, then
\[ \sum^{N-1}_{\alpha=0} (\mu|\hat{\cal E}^{\dagger}|\alpha)(\alpha|
\hat{\cal E}|\nu)=(\mu|\hat I|\nu), \]
i.e.
\[ \sum^{N-1}_{\alpha=0} ({\cal E}^{\dagger})_{\mu \alpha}
{\cal E}_{\alpha \nu}=\delta_{\mu \nu}. \]
Using proposition 9 we have
\[ \sum^{N-1}_{\alpha=0} (\hat{\cal E}^{T})_{\mu \alpha}
{\cal E}_{\alpha \nu}=\delta_{\mu \nu}, \]
i.e. ${\cal E}^{T}{\cal E}=I$.

If ${\cal E}_{\mu \nu}$ is real orthogonal matrix, then
\[ \sum^{N-1}_{\nu=0} ({\cal E}_{\mu \nu})^{2}=1. \]
Therefore all elements of orthogonal gate matrix never exceed 1,
i.e. $|{\cal E}_{\mu \nu}| \le 1$.

Note that n-qubit unitary two-valued logic gate
$U$ is an element of Lie group $SU(2^{n})$.
The dimension of this group is equal to
$dim \ SU(2^{n})=(2^{n})^{2}-1=4^{n}-1$.
The matrix of n-ququat orthogonal linear gate
$\hat{\cal U}=\hat{\cal E}^{(U)}$ can be considered as an element
of Lie group $SO(4^{n}-1)$.
The dimension of this group is equal to
$dim \ SO(4^{n}-1)=(4^{n}-1)(2 \cdot 4^{n-1}-1)$.

For example, if $n=1$, then
\[ dim \ SU(2^{1})=3 \ , \quad dim \  SO(4^{1}-1)=3. \]
If $n=2$, then
\[ dim \ SU(2^{2})=15 \ , \quad dim \  SO(4^{2}-1)=105. \]
Therefore not all orthogonal 4-valued logic gates for mixed
and pure states are connected
with unitary 2-valued logic gates for pure states.

\subsection{Single ququat orthogonal gates}

Let us consider single ququat 4-valued logic gate $\hat{\cal U}$
associated with unitary single qubit 2-valued logic gate $U$.


\bp {\it Any single-qubit unitary quantum two-valued logic gate
can be realized as the product of single ququat simple rotation
gates $\hat{\cal U}^{(1)}(\alpha)$, $\hat{\cal U}^{(2)}(\theta)$
and $\hat{\cal U}^{(1)}(\beta)$ defined by
\[ \hat{\cal U}^{(1)}(\alpha)=|0)(0|+|3)(3|+
\cos \alpha \Bigl(|1)(1|+|2)(2| \Bigr)+\]
\[+\sin \alpha \Bigl(|2)(1|-|1)(2| \Bigr), \]
\[ \hat{\cal U}^{(2)}(\theta)=|0)(0|+|2)(2|+
\cos \theta \Bigl(|1)(1|+|3)(3| \Bigr)+\]
\[+\sin \theta \Bigl(|1)(3|-|3)(1| \Bigr), \]
where $\alpha$, $\theta$ and $\beta$ are Euler angles.
}
\ep

\noindent {\bf Proof.}
Let us consider a general single qubit unitary gate \cite{Bar}.
Every unitary one-qubit gate $U$ can be represented by $2 \times 2$-matrix
\[  U(\alpha,\theta,\beta)=e^{-i \alpha \sigma_{3}/2}
e^{-i \theta \sigma_{2}/2}e^{-i \beta\sigma_{3}/2}= \]
$$=\left(
\begin{array}{rr}
e^{-i (\alpha/2 + \beta/2)} \cos \theta/2 &
-e^{-i (\alpha/2 - \beta/2)} \sin \theta/2 \\
e^{i (\alpha/2 - \beta/2)} \sin \theta/2 &
e^{i (\alpha/2 + \beta/2)} \cos \theta/2
\end{array}
\right),$$
i.e.
$$
U(\alpha,\theta,\beta)=U_1(\alpha)U_2(\theta)U_1(\beta),$$
where
$$U_1(\alpha)=\left(
\begin{array}{ll}
e^{-i \alpha/2} & 0 \\
0        & e^{i \alpha/2}
\end{array}
\right),$$
$$U_2(\theta)=\left(
\begin{array}{rr}
\cos \theta/2 &- \sin \theta/2 \\
\sin \theta/2 & \cos \theta/2
\end{array}
\right),$$
$$
U_1(\beta)=\left(
\begin{array}{ll}
e^{-i \beta/2} & 0 \\
0        & e^{i \beta/2}
\end{array}
\right),
$$
where $\alpha$, $\theta$ and $\beta$ are Euler angles.
The correspondent $4\times 4$-matrix ${\cal U}(\alpha,\theta,\beta)$
of four-valued logic gate has the form
\[ {\cal U}(\alpha,\theta,\beta)={\cal U}^{(1)}(\alpha)
{\cal U}^{(2)}(\theta){\cal U}^{(1)}(\beta), \]
where
\[ {\cal U}^{(1)}_{\mu \nu}(\alpha)=
\frac{1}{2}Tr\Bigl(\sigma_{\mu}
U_1(\alpha)\sigma_{\nu} U^{\dagger}_1(\alpha)\Bigr), \]
\[ {\cal U}^{(2)}_{\mu \nu}(\theta)=
\frac{1}{2}Tr\Bigl(\sigma_{\mu}
U_2(\theta)\sigma_{\nu} U^{\dagger}_2(\theta)\Bigr), \]
Finally, we obtain
$${\cal U}^{(1)}(\alpha)=\left(
\begin{array}{cccc}
1&0&0&0\\
0&\cos \alpha & -\sin \alpha & 0 \\
0&\sin \alpha & \cos \alpha & 0 \\
0&0 & 0 &  1
\end{array}
\right),$$
$${\cal U}^{(2)}(\theta)=\left(
\begin{array}{cccc}
1&0&0&0\\
0&\cos \theta & 0 & \sin \theta \\
0&0 & 1 & 0 \\
0&-\sin \theta & 0 & \cos \theta
\end{array}
\right),$$
where
\[ 0  \le \alpha <2 \pi , \quad  0\le \theta \le \pi ,
\quad 0 \le \beta\le 2\pi. \]
Using $U(\alpha,\theta +2\pi, \beta)=- U(\alpha,\theta, \beta)$,
we get that 2-valued logic gates $U(\alpha,\theta, \beta)$ and
$U(\alpha,\theta +2\pi, \beta)$ map into single
4-valued logic gate ${\cal U}(\alpha,\theta, \beta)$.
The back rotation 4-valued logic gate is defined by the matrix
\[ {\cal U}^{-1}(\alpha,\theta, \beta)=
{\cal U}(2\pi -\alpha, \pi- \theta,2\pi - \beta)  \ . \]
The simple rotation gates
$\hat{\cal U}^{(1)}(\alpha)$, $\hat{\cal U}^{(2)}(\theta)$,
$\hat{\cal U}^{(1)}(\beta)$
are defined by matrices $\hat{\cal U}^{(1)}(\alpha)$,
$\hat{\cal U}^{(2)}(\theta)$ and $\hat{\cal U}^{(1)}(\beta)$.

Let us introduce simple reflection gates by
\[ \hat{\cal R}^{(1)}=|0)(0|-|1)(1|+|2)(2|+|3)(3|, \]
\[ \hat{\cal R}^{(2)}=|0)(0|+|1)(1|-|2)(2|+|3)(3|, \]
\[ \hat{\cal R}^{(3)}=|0)(0|+|1)(1|+|2)(2|-|3)(3|. \]

\bp
{\it Any single ququat linear gate $\hat{\cal E}$
defined by orthogonal matrix ${\cal E}:$ \ ${\cal E}{\cal E}^T=I$
can be realized by
\begin{itemize}
\item simple rotation gates $\hat{\cal U}^{(1)}$ and
$\hat{\cal U}^{(2)}$.
\item inversion gate $\hat {\cal I}$ defined by
\[ \hat{\cal I}=|0)(0|-|1)(1|-|2)(2|-|3)(3|. \]
\end{itemize}
}
\ep

\noindent {\bf Proof.} Using Proposition 11 and
\[ \hat{\cal R}^{(3)}=\hat{\cal U}^{(1)} \hat{\cal I}, \quad
\hat{\cal R}^{(2)}=\hat{\cal U}^{(2)} \hat{\cal I} \ , \quad
\hat{\cal R}^{(1)}=\hat{\cal U}^{(1)}
\hat{\cal U}^{(1)} \hat{\cal I}, \]
we get this proposition.

\noindent{\bf Example 1.}
In the generalized computational basis the Pauli matrices
as two-valued logic gates are the four-valued logic gates
with diagonal $4\times 4$ matrix. The gate $I=\sigma_0$ is
\[ \hat{\cal U}^{(\sigma 0)}=\sum^3_{\mu=0}|\mu)(\mu|=\hat I, \]
i.e.
${\cal U}^{(\sigma 0)}_{\mu \nu}=
(1/2)Tr(\sigma_{\mu}\sigma_{\nu})=\delta_{\mu \nu}$.

For the unitary two-valued logic gates are equal to the Pauli matrix
$\sigma_k$, where $k \in \{1,2,3\}$,
we have quantum four-valued logic gates
\[ \hat{\cal U}^{(\sigma k)}=\sum^3_{\mu , \nu=0}
{\cal U}^{(\sigma k)}_{\mu \nu} \ |\mu)(\nu|, \]
with the matrix
\begin{equation} \label{9}
{\cal U}^{(\sigma k)}_{\mu \nu}=2 \delta_{\mu 0} \delta_{\nu 0}+
2 \delta_{\mu k} \delta_{\nu k}- \delta_{\mu \nu} \ .
\end{equation}

\noindent{\bf Example 2.}
In the generalized computational basis the unitary NOT gate
("negation") of two-valued logic
\[ X=|0><1|+|1><0|=\sigma_{1}=
\left(
\begin{array}{cc}
0&1\\
1&0\\
\end{array}
\right), \]
is represented by quantum four-valued logic gate
\[ \hat{\cal U}^{(X)}=|0)(0|+|1)(1|-|2)(2|-|3)(3|, \]
i.e.  $4 \times 4$ matrix is
$${\cal U}^{(X)}=\left(
\begin{array}{cccc}
1&0&0&0\\
0&1&0&0\\
0&0&-1&0\\
0&0&0&-1
\end{array}
\right).$$

\noindent{\bf Example 3.}
The Hadamar two-valued logic gate
\[ H=\frac{1}{\sqrt{2}}(\sigma_1+\sigma_3)  \]
can be represented as a four-valued logic gate by
\[ \hat{\cal E}^{(H)}=|0)(0|-|2)(2|+|3)(1|+|1)(3|, \]
with
\[ {\cal E}^{(H)}_{\mu \nu}=\delta_{\mu 0} \delta_{\nu 0}-
\delta_{\mu 2} \delta_{\nu 2}+\delta_{\mu 3} \delta_{\nu 1}+
\delta_{\mu 1} \delta_{\nu 3}. \]

\subsection{Measurements as quantum 4-valued logic gates}

It is known that superoperator $\hat{\cal E}$ of von Neumann measurement is
defined by
\begin{equation} \label{vNm}
\hat {\cal E}|\rho)=\sum^{r}_{k=1} |P_{k}\rho P_{k}) \ , \end{equation}
where $\{P_{k}|k=1,..,r\}$ is a (not necessarily complete)
sequence of orthogonal projection operators on ${\cal H}^{(n)}$.

Let $P_{k}$ are projectors onto the pure state $|k>$
which define usual computational basis $\{|k>\}$, i.e.
$P_{k}=|k><k|$.

\bp
{\it A nonlinear four-valued logic gate $\hat{\cal N}$
for von Neumann measurement (\ref{vNm}) 
of the state $\rho=\sum^{N-1}_{\alpha=0}|\alpha) \rho_{\alpha}$
is defined by
\[ \hat{\cal N}=\sum^{r}_{k=1} \frac{1}{p(k)}
{\cal E}^{(k)}_{\mu \nu} |\mu)(\nu|, \]
where
\begin{equation} \label{EE1}
{\cal E}^{(k)}_{\mu \nu}=
\frac{1}{2^{n}} Tr(\sigma_{\mu} P_{k} \sigma_{\nu} P_{k}),
\quad p(k)=\sqrt{2^n} \sum^{N-1}_{\alpha=0} {\cal E}^{(k)}_{0 \alpha}
\rho_{\alpha} \ . \end{equation}
}
\ep

\noindent {\bf Proof.}
The trace-decreasing superoperator $\hat{\cal E}_{k}$ is defined by
\[ |\rho) \ \rightarrow \ |\rho^{\prime})={\cal E}_{k}|\rho)=|P_{k}\rho P_{k}). \]
The superoperator $\hat{\cal E}$ has
the form $\hat{\cal E}_{k}=\hat L_{P_{k}} \hat R_{P_{k}}$.
Then
\[ {\rho'}_{\mu}=(\mu|\rho')=(\mu|\hat{\cal E}_{k}|\rho)=
\sum^{N-1}_{\nu=0} (\mu|\hat{\cal E}_{k}|\nu)(\nu|\rho)=
\sum^{N-1}_{\nu=0} {\cal E}^{(k)}_{\mu \nu} \rho_{\nu}\ , \]
where
\[ {\cal E}^{(k)}_{\mu \nu}=(\mu|\hat{\cal E}_{k}|\nu)=
\frac{1}{2^{n}} Tr(\sigma_{\mu} P_{k} \sigma_{\nu} P_{k}). \]
The probability that process represented by $\hat{\cal E}_{k}$ 
occurs is
\[ p(k)=Tr(\hat{\cal E}_k(\rho))=(I|\hat{\cal E}|\rho)=
\sqrt{2^n}\rho^{\prime}_0=\sqrt{2^n}
\sum^{N-1}_{\alpha=0} {\cal E}^{(k)}_{0 \alpha} \rho_{\alpha} \ . \]
If
\[ \sum^{N-1}_{\alpha=0} {\cal E}_{0 \alpha} \rho_{\alpha} \not=0 \ ,
\]
then the matrix for nonlinear trace-preserving gate $\hat{\cal N}$ is
\[ {\cal N}_{\mu \nu}=\sqrt{2^n}
(\sum^{N-1}_{\alpha=0}{\cal E}_{0 \alpha} \rho_{\alpha})^{-1}
{\cal E}_{\mu \nu}. \]

\noindent{\bf Example.}
Let us consider single ququat projection operator
\[ P_{0}=|0><0|=\frac{1}{2}(\sigma_{0}+\sigma_{3}). \]
Using formula (\ref{EE1}) we derive
\[ {\cal E}^{(0)}_{\mu \nu}=
\frac{1}{8}Tr\Bigl(\sigma_{\mu}(\sigma_{0}+\sigma_{3})
\sigma_{\nu}(\sigma_{0}+\sigma_{3}) \Bigr)=\]
\[=\frac{1}{2}\Bigl(
\delta_{\mu 0}\delta_{\nu 0}+\delta_{\mu 3} \delta_{\nu 3}+
\delta_{\mu 3}\delta_{\nu 0}+\delta_{\mu 0} \delta_{\nu 3}
\Bigr), \]
i.e.
$${\cal E}^{(0)}=\left(
\begin{array}{cccc}
1/2&0&0&1/2\\
0&0&0&0\\
0&0&0&0\\
1/2&0&0&1/2
\end{array}
\right).$$
The linear trace-decreasing superoperator
for von Neumann measurement
projector $|0><0|$ onto pure state $|0>$ is
\[ \hat{\cal E}^{(0)}=\frac{1}{2}\Bigl( |0)(0)+|3)(3|+
|0)(3|+|3)(0| \Bigr). \]

\noindent{\bf Example.}
For the projection operator
\[ P_{1}=|1><1|=\frac{1}{2}(\sigma_{0}-\sigma_{3}) \]
Using formula (\ref{EE1}) we derive
\[ {\cal E}^{(1)}_{\mu \nu}=
\frac{1}{2}\Bigl(
\delta_{\mu 0}\delta_{\nu 0}+\delta_{\mu 3} \delta_{\nu 3}-
\delta_{\mu 3}\delta_{\nu 0}-\delta_{\mu 0} \delta_{\nu 3}
\Bigr). \]
The linear superoperator $\hat{\cal E}^{(1)}$ for von Neumann measurement
projector onto pure state $|1>$ is
\[ \hat{\cal E}^{(1)}=
\frac{1}{2}\Bigl( |0)(0)+|3)(3|-|0)(3|-|3)(0| \Bigr), \]
i.e.
$${\cal E}^{(1)}=\left(
\begin{array}{cccc}
1/2&0&0&-1/2\\
0&0&0&0\\
0&0&0&0\\
-1/2&0&0&1/2
\end{array}
\right).$$

The superoperators $\hat{\cal E}^{(0)}$ and $\hat{\cal E}^{(1)}$
are not trace-preserving.
The probabilities that processes represented by superoperators
$\hat{\cal E}^{(k)}$ occurs are
\[ p(0)=\frac{1}{\sqrt{2}}(\rho_0+\rho_{3}) \ , \quad
p(1)=\frac{1}{\sqrt{2}}(\rho_0+\rho_{3}). \]



\subsection{Reversible quantum 4-valued logic gate}

In the paper \cite{MZ}, Mabuchi and Zoller have shown how a measurement
on a quantum system can be reversed under appropriate conditions.
In the papers \cite{KL,NCSB,NC,BDSW} was considered necessary
and sufficient conditions for general quantum operations
to be reversible.

Let us consider quantum operation ${\cal E}$
on a subspace ${\cal M}$ of the total state space.

\noindent {\bf Theorem 6.}
{\it A quantum operation ${\cal E}$
\[ {\cal E}(\rho)=\sum^{m}_{j=1} A_{j} \rho A^{\dagger}_{j} \]
is reversible on subspace ${\cal M}$ if and only if there exists
a positive matrix $M$ such that
\begin{equation} \label{PAAP}
P_{\cal M}A^{\dagger}_{k}A_{j}P_{\cal M}=M_{jk}P_{\cal M} \ .
\end{equation}
where $P_{\cal M}$ is a projector onto subspace ${\cal M}$.
The trace of $M$
\[ \sum^{m}_{j=1} M_{jj}=\mu^2 \]
is the constant value of \ $Tr({\cal E}(\rho))$ on ${\cal M}$.}

\noindent {\bf Proof.} This result was proved in \cite{BDSW,KL,NCSB}.

Let $\hat{\cal E}^{({\cal M})}$ is  projection superoperator defined by
\[ \hat{\cal E}^{\cal M}(\rho)=P_{\cal M} \rho P_{\cal M}. \]
Note that
\[ \hat{\cal E}^{({\cal M})}\hat{\cal E}^{({\cal M})}=
\hat{\cal E}^{({\cal M})} \ , \quad
(\hat{\cal E}^{({\cal M})})^{\dagger}=\hat{\cal E}^{({\cal M})}. \]

Let $\hat{\cal E}_{\cal M}$ be the restriction of $\hat{\cal E}$ to the
subspace ${\cal M}$
\begin{equation} \label{ED} \hat {\cal E}_{\cal M}(\rho)=\sum^{m}_{j=1}
A_{j}P_{\cal M}\rho P_{\cal M} A^{\dagger}_{j} \ . \end{equation}
Notice that $\hat {\cal E}_{\cal M}(\rho)=\hat {\cal E}(\rho)$ if
$\rho$ lies wholly in ${\cal M}$.
Note, that the adjoint superoperator for trace-decreasing quantum
operation is generally not a quantum operation, since it can be
trace-increasing, but it is always a completely positive map.

Equation (\ref{PAAP}) is equivalent to the requirement that
superoperator ${\cal E}^{\dagger}_{\cal M}(\rho)$ be a positive
multiple of identity operation on ${\cal M}$.
This requirement can be formulated as theorem.

\noindent {\bf Theorem 7.}\\
{\it A necessary and sufficient condition for reversibility
of linear superoperator $\hat{\cal E}$ on the subspace ${\cal M}$ is
\[ \hat{\cal E}^{\cal M} \hat{\cal E}^{\dagger} \hat{\cal E}
\hat{\cal E}^{\cal M}= \gamma \hat{\cal E}^{\cal M}. \]
}

\noindent {\bf Proof.} For the proofs we refer to \cite{NCSB}.


\section{Classical four-valued logic classical gates}


Let us consider some elements of classical four-valued logic.
For the concept of many-valued logic see \cite{Re,RT,Ya1,Ya2,Ta}.

\subsection{Elementary classical gates}

A classical four-valued logic gate is called a function
$g(x_{1},...,x_{n})$ if following conditions hold:
\begin{itemize}
\item all $x_{i} \in \{0,1,2,3\}$, where $i=1,...,n$.
\item $g(x_{1},...,x_{n}) \in \{0,1,2,3\}$.
\end{itemize}

It is known that
the number of all classical logic gates with n-arguments
$x_{1},...,x_{n}$ is equal to
$4^{4^{n}}$.
The number of  classical logic gates $g(x)$ with single argument
is equal to $4^{4^{1}}=256$.

\vskip 5mm
\begin{tabular}{|c|c|c|c|c|c|c|c|c|}
\hline
\multicolumn{9}{|c|}{Single argument classical gates}\\
\hline
x& $\sim x$ &$\Box x$&$\diamondsuit x$&$\overline{x}$&$I_{0}$&$I_{1}$&$I_{2}$&$I_{3}$\\ \hline
0& 3        &   0    &      0         &      1       &  3    &   0   &   0   & 0   \\
1& 2        &   0    &      3         &      2       &  0    &   3   &   0   & 0    \\
2& 1        &   0    &      3         &      3       &  0    &   0   &   3   & 0   \\
3& 0        &   3    &      3         &      0       &  0    &   0   &   0   & 3    \\ \hline
\end{tabular}
\vskip 3mm

\begin{tabular}{|c|c|c|c|c|c|c|c|}
\hline
\multicolumn{8}{|c|}{Single argument classical gates}\\
\hline
x&0&1&2&3&$g_1$&$g_2$&$g_{3}$\\
\hline
0&0&1&2&3&  3  &  0  &   1   \\
1&0&1&2&3&  0  &  1  &   1   \\
2&0&1&2&3&  1  &  3  &   2   \\
3&0&1&2&3&  2  &  2  &   3   \\ \hline
\end{tabular}
\vskip 3mm

The number of  classical logic gates $g(x_{1},x_{2})$ with
two-arguments is equal to
\[ 4^{4^{2}}=4^{16}=42949677296. \]
Let us write some of these gates.

\vskip 5mm
\begin{tabular}{|c|c|c|c|c|}
\hline
\multicolumn{5}{|c|}{Two-arguments classical gates}\\
\hline
$(x_{1},x_{2})$&$\land$&$\lor$&$V_{4}$&$\sim V_{4}$ \\ \hline
(0;0)          &   0   &   0  &  1    &    2        \\
(0;1)          &   0   &   1  &  2    &    1        \\
(0;2)          &   0   &   2  &  3    &    0        \\
(0;3)          &   0   &   3  &  0    &    3        \\
(1;0)          &   0   &   1  &  2    &    1        \\
(1;1)          &   1   &   1  &  2    &    1        \\
(1;2)          &   1   &   2  &  3    &    0        \\
(1;3)          &   1   &   3  &  0    &    3        \\
(2;0)          &   0   &   2  &  3    &    0        \\
(2;1)          &   1   &   2  &  3    &    0        \\
(2;2)          &   2   &   2  &  3    &    0        \\
(2;3)          &   2   &   3  &  0    &    3        \\
(3;0)          &   0   &   3  &  0    &    3        \\
(3;1)          &   1   &   3  &  0    &    3        \\
(3;2)          &   2   &   3  &  0    &    3        \\
(3;3)          &   3   &   3  &  0    &    3        \\ \hline
\end{tabular}
\vskip 3mm

Let us define some elementary classical 4-valued logic gates by formulas.
\begin{itemize}
\item Luckasiewicz negation: \  $\sim x=3-x$.
\item Cyclic shift: \ $\overline{x}=x+1(mod4)$.
\item Functions $I_{i}(x)$, where $i=0,...,3$, such that
$I_{i}(x)=3$ if $x=i$ and $I_{i}(x)=0$ if $x\not=i$.
\item Generalized conjunction: \ $x_{1} \land x_{2}=min(x_{1},x_{2})$.
\item Generalized disjunction: \ $x_{1} \lor x_{2}=max(x_{1},x_{2})$.
\item Generalized Sheffer function: \
$$V_{4}(x_{1},x_{2})=max(x_{1},x_{2})+1(mod 4).$$
\end{itemize}

Commutative law, associative law and distributive law for the
generalized conjunction and disjunction are satisfied:
\begin{itemize}
\item Commutative law
\[ x_{1} \land x_{2}=x_{2} \land x_{1} \ , \quad
x_{1} \lor x_{2}=x_{2} \lor x_{1} \]
\item Associative law
\[ (x_{1} \lor x_{2}) \lor x_{3}=
x_{1} \lor (x_{2} \lor x_{3}). \]
\[ (x_{1} \land x_{2}) \land x_{3}=
x_{1} \land (x_{2} \land x_{3}). \]
\item Distributive law
\[ x_{1}\lor (x_{2} \land x_{3})=
(x_{1} \lor x_{2}) \land ( x_{1} \lor x_{3}). \]
\[ x_{1}\land (x_{2} \lor x_{3})=
(x_{1} \land x_{2}) \lor ( x_{1} \land x_{3}). \]
\end{itemize}
Note that the Luckasiewicz negation is satisfied
\[ \sim(\sim x)=x \ , \quad
\sim(x_{1} \land x_{2})=(\sim x_{1}) \lor (\sim x_{2}). \]
The shift $\overline{x}$ for $x$ is not satisfied usual negation rules:
\[ \overline{\overline{x}} \not=x \ , \quad \overline{x_{1}\land x_{2}}
\not=\overline{x_{1}} \lor \overline{x_{2}}. \] The analog of
disjunction normal form of the n-arguments 4-valued logic gate is
\[ g(x_{1},...,x_{n})=\bigvee_{(k_{1},...,k_{n})} \ I_{k_{1}}(x_{1})
\land ... \land I_{k_{n}} \land g(k_{1},..,k_{n}). \]

\subsection{Universal classical gates}

Let us consider universal sets of universal classical gates of four-valued
logic.

\noindent {\bf Theorem 8.}\\
{\it
The set $\{0, 1, 2, 3, I_{0}, I_{1}, I_{2}, I_{3},
x_{1} \land x_{2}, x_{1} \lor x_{2}\}$ is universal.\\
The set $\{\overline{x}, x_{1} \lor x_{2}\}$ is universal.\\
The gate $V_{4}(x_{1}, x_{2})$ is universal.
}

\noindent {\bf Proof.} This theorem is proved in \cite{Ya2}.

\noindent {\bf Theorem 9.} \\
{\it All logic single argument 4-valued gates $g(x)$
can be generated by functions:
\begin{itemize}
\item $g_{1}(x)=x-1(mod4)$.
\item $g_{2}(x)$: $g_{2}(0)=0$, $g_{2}(1)=1$, $g_{2}(2)=3$, $g_{2}(3)=2$.
\item $g_{3}(x)=1$ if $x=0$ and $g_{3}(x)=x$ if $x\not=0$ .
\end{itemize}
} \noindent {\bf Proof.} This theorem was proved by Piccard in
\cite{Pic}.


\section{Quantum four-valued logic gates for classical gates}


\subsection{Quantum gates for single argument classical gates}

Let us consider linear trace-preserving quantum gates for
classical gates $\sim$, $\overline{x}$,
$I_{0},I_{1},I_{2},I_{3}$, $0,1,2,3$,
$g_{1},g_{2},g_{3}$,  $\diamondsuit$, $\Box$.

\bp
{\it Any single argument classical gate $g(\nu)$
can be realized as linear trace-preserving quantum
four-valued logic gate by
\[ \hat{\cal E}^{(g)}=|0)(0|+\sum^{3}_{k=1 }|g(k))(k|+\]
\[+(1-\delta_{0g(0) })\Bigl( |g(0))(0|-
\sum^{3}_{\mu=0} \sum^{3}_{\nu=0}(1-\delta_{\mu g(\nu)})
|\mu)(\nu| \Bigr). \]
}
\ep

\noindent {\bf Proof.}
The proof is by direct calculation in
\[ \hat{\cal E}^{(g)}|\alpha]=|g(\alpha)], \]
where
\[ \hat{\cal E}^{(g)}|\alpha]= \frac{1}{\sqrt{2}}\Bigl(
\hat{\cal E}^{(g)}|0)+\hat{\cal E}^{(g)}|\alpha) \Bigr). \]

\noindent{\bf Examples.}

1. Luckasiewicz negation gate is
\[ \hat{\cal E}^{(LN)}=|0)(0|+|1)(2|+|2)(1|+|3)(0|-|3)(3|. \]
$${\cal E}^{(LN)}=\left(
\begin{array}{cccc}
1&0&0&0\\
0&0&1&0\\
0&1&0&0\\
1&0&0&-1
\end{array}
\right).$$

2.  The four-valued logic gate $I_{0}$ can be realized by
\[ \hat{\cal E}^{(I_0)}=|0)(0|+|3)(0|-\sum^{3}_{k=1}|3)(k|. \]
$${\cal E}^{(I_{0})}=\left(
\begin{array}{cccc}
1&0&0&0\\
0&0&0&0\\
0&0&0&0\\
1&-1&-1&-1
\end{array}
\right).$$

3. The gates $I_{k}(x)$, where $k=1,2,3$ is
\[ \hat{\cal E}^{(I_k)}=|0)(0|+|3)(k|. \]
For example,
$${\cal E}^{(I_{1})}=\left(
\begin{array}{cccc}
1&0&0&0\\
0&0&0&0\\
0&0&0&0\\
0&1&0&0
\end{array}
\right),$$

4.  The gate $\overline{x}$ can be realized by
\[ \hat{\cal E}^{(\overline{x})}=|0)(0|+|1)(0|+|2)(1|+|3)(2|-
\sum^{3}_{k=1}|1)(k|. \]
$${\cal E}^{(\overline{x})}=\left(
\begin{array}{cccc}
1&0&0&0\\
1&-1&-1&-1\\
0&1&0&0\\
0&0&1&0
\end{array}
\right).$$

5. The constant gates $0$ and $k=1,2,3$ can be realized by
\[ \hat{\cal E}^{(0)}=|0)(0|  \ , \quad
\hat{\cal E}^{(k)}=|0)(0|+|k)(0|. \]
For example,
$${\cal E}^{(1)}=\left(
\begin{array}{cccc}
1&0&0&0\\
1&0&0&0\\
0&0&0&0\\
0&0&0&0
\end{array}
\right).$$

6.  The gate $g_{1}(x)$ can be realized by
\[ \hat{\cal E}^{(g_{1})}=|0)(0|+|1)(2|+|2)(3|+|3)(0|-
\sum^{3}_{k=1}|3)(k|. \]
$${\cal E}^{(g_{1})}=\left(
\begin{array}{cccc}
1&0&0&0\\
0&0&1&0\\
0&0&0&1\\
1&-1&-1&-1
\end{array}
\right).$$

7.  The gate $g_{2}(x)$ is
\[ \hat{\cal E}^{(g_{2})}=|0)(0|+|1)(1|+|3)(2|+|2)(3|. \]
$${\cal E}^{(g_{2})}=\left(
\begin{array}{cccc}
1&0&0&0\\
0&1&0&0\\
0&0&0&1\\
0&0&1&0
\end{array}
\right).$$

8.  The gate $g_{3}(x)$ can be realized by
\[ \hat{\cal E}^{(g_{3})}=|0)(0|+|2)(2|+|3)(3|+ |1)(0|-|1)(2|-|1)(3|. \]
$${\cal E}^{(g_{3})}=\left(
\begin{array}{cccc}
1&0&0&0\\
1&0&-1&-1\\
0&0&1&0\\
0&0&0&1
\end{array}
\right).$$

9.  The gate $\diamondsuit x$ is
\[ \hat{\cal E}^{(\diamondsuit)}=|0)(0|+\sum^{3}_{k=1}|3)(k|. \]
$${\cal E}^{(\diamondsuit)}=\left(
\begin{array}{cccc}
1&0&0&0\\
0&0&0&0\\
0&0&0&0\\
0&1&1&1
\end{array}
\right).$$

10. The gate $\Box x= \sim \diamondsuit x$ is
\[ \hat{\cal E}^{(\sim \diamondsuit)}=|0)(0|+|3)(3|. \]
$${\cal E}^{(\Box)}=\left(
\begin{array}{cccc}
1&0&0&0\\
0&0&0&0\\
0&0&0&0\\
0&0&0&1
\end{array}
\right).$$

Note that quantum gates $\hat{\cal E}^{(LN)}$,
$\hat{\cal E}^{(I_{0})}$, $\hat{\cal E}^{(k)}$,
$\hat{\cal E}^{(g_{1})}$ are not unital gates.

\subsection{Quantum gates for two-arguments classical gates}

Let us consider quantum gates for two-arguments classical gates.

1. The generalized conjunction $x_{1} \land x_{2}=min(x_{1},x_{2})$ and
generalized disjunction $x_{1} \land x_{2}=max(x_{1},x_{2})$ can be
realized by two-ququat gate with $T=0$:
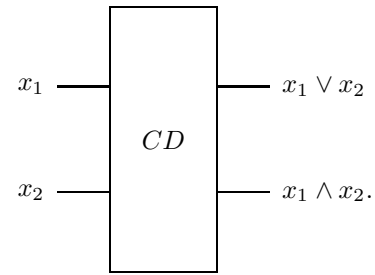
\begin{center}
\begin{picture}(10,10)(0,0)
\put(3,0){\framebox(4,10){$CD$}}
\put(1,3){\line(1,0){2}}
\put(1,7){\line(1,0){2}}
\put(7,3){\line(1,0){2}}
\put(7,7){\line(1,0){2}}
\put(0.5,7){\makebox(0,0)[r]{$x_{1}$}}
\put(0.5,3){\makebox(0,0)[r]{$x_{2}$}}
\put(9.5,7){\makebox(0,0)[l]{$x_{1} \lor x_{2}$}}
\put(9.5,3){\makebox(0,0)[l]{$x_{1} \land x_{2}.$}}
\end{picture}
\end{center}

Let us write the quantum gate
which realizes the CD gate in the generalized computational basis by
\[ \hat{\cal E}=\sum^{N-1}_{\mu}  \sum^{N-1}_{\nu} |\mu\nu)(\mu \nu|+
\sum^{3}_{k=1}\Bigl( |0k)-|k0) \Bigr)(k 0|+\]
\[ +\sum^{3}_{k=2}\Bigl( |1k)-|k1) \Bigr)(k 1|+
\Bigl( |23)-|32) \Bigl)(32|. \]

2.  The Sheffer function gate $|x_{1},x_{2}] \  \rightarrow \
|V_{4}(x_{1},x_{2}),\sim V_{4}(x_{1},x_{2})]$ can be realized
by  two-ququat gate with $T\not=0$:
\[ \hat{\cal E}^{(SF)}=|00)(00|+|12)(00|-
\sum^{3}_{\mu=0}\sum^{3}_{\nu=1}|12)(\mu \nu|+|21)(10|+ \]
\[ +|21)(11|+|30)(02|+|30)(20|+|30)(12|+|30)(21|+\]
\[ +|30)(22|+|03)(03|+|03)(13|+|03)(23|+\sum^{3}_{\mu=0} |03)(3\mu|. \]

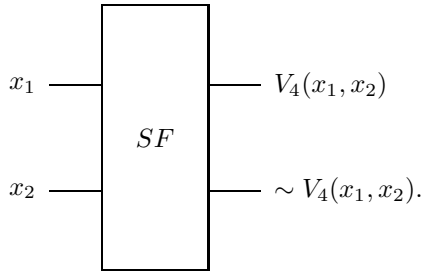
\begin{center}
\begin{picture}(10,10)(0,0)
\put(3,0){\framebox(4,10){$SF$}}
\put(1,3){\line(1,0){2}}
\put(1,7){\line(1,0){2}}
\put(7,3){\line(1,0){2}}
\put(7,7){\line(1,0){2}}
\put(0.5,7){\makebox(0,0)[r]{$x_{1}$}}
\put(0.5,3){\makebox(0,0)[r]{$x_{2}$}}
\put(9.5,7){\makebox(0,0)[l]{$V_{4}(x_{1},x_{2})$}}
\put(9.5,3){\makebox(0,0)[l]{$\sim V_{4}(x_{1},x_{2}).$}}
\end{picture}
\end{center}

Note that this Sheffer function gate is not unital quantum gate and
\[ \hat{\cal E}^{(SF)} \not=
|V_{4}(x_{1},x_{2}),\sim V_{4}(x_{1},x_{2}))(x_{1},x_{2}|. \]

\subsection{Unital quantum gates for single argument classical gates}

It is interesting to consider a representation for classical gates by
linear unital quantum gates ($\hat{\cal E}|0]=|0]$).
There is a restriction for representation
single argument classical (4-valued logic) gate
by linear quantum four-valued logic gates with $T=0$ (all $T_k=0$).
Any unital n-ququat quantum gate has the form
\[ \hat{\cal E}^{(T=0)}=|0)(0|+\sum^{N-1}_{\mu=1}
\sum^{N-1}_{\nu=1} R_{\mu \nu}|\mu)(\nu|, \]
i.e. $T_k=0$ for all $k$ and the gate matrix is
$${\cal E}^{(T=0)}=\left(
\begin{array}{ccccc}
1&0&0&...&0\\
0&R_{11}&R_{12}&...&R_{1 \ N-1}\\
0&R_{21}&R_{22}&...&R_{2 \ N-1}\\
...&...&...& ...&...\\
0&R_{N-1 \ 1}&R_{N-1 \ 2}&...&R_{N-1 \ N-1}
\end{array}
\right).$$


\bp
{\it If the single argument classical (4-valued logic) gate
$g(x)$ such that $g(0)=k$, where $k \in\{1,2,3\}$
and exists $m \in\{1,2,3\}$: $g(m)=l$, where $l\not=k$,  then
there is no a representation of this gate by some unital quantum
four-valued logic gates $\hat{\cal E}$.}
\ep

\noindent {\bf Proof.}
If $\hat{\cal E}|0]=|k]$ and $\hat{\cal E}|m]=|l]$ ,
where $k \in\{1,2,3\}$, $l\not=k$,
then
\[ {\cal E}_{k0}=(k|\hat{\cal E}|0)=(k| \hat{\cal E}|0]=(k|k]=\]
\[=\frac{1}{\sqrt{2^n}} \Bigl( (k|0)+(k|k) \Bigr)=
\frac{1}{\sqrt{2^n}}\not=0, \]
i.e. $T_{k}\not=0$ in the matrix ${\cal E}_{\mu \nu}$.


From this proposition we see that single argument classical gate
$g(x)$ can be realized by single ququat quantum gate with $T=0$
if and only if \\
1. $g(0)=0$, or\\
2. $g(\mu)=const$, i.e. $g(0)=g(1)=g(2)=g(3)=k$ and
$k \in \{1,2,3\}$.\\
For example, classical gates
$ \sim x$, $I_{0}$, $\overline{x}$, $g_{1}$ and $g_{3}$
can not be realized by single ququat unital quantum gates.


Single argument classical logic gates such that $g(0)\not=0$ can not
be realized by single ququat  quantum gates $\hat{\cal E}$ with $T=0$.
This classical gates can be realized
by two-qubits unital quantum gates.
Let us consider Luckasiewicz negation $\sim x=3-x$.
If $x_{2}\not=0$ and $x_{1}, x_{2} \in \{0,1,2,3\}$ then
we can define quantum Luckasiewicz negation gate by
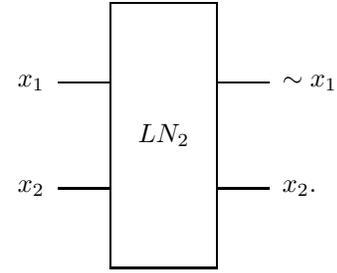
\begin{center}
\begin{picture}(10,10)(0,0)
\put(3,0){\framebox(4,10){$LN_{2}$}}
\put(1,3){\line(1,0){2}}
\put(1,7){\line(1,0){2}}
\put(7,3){\line(1,0){2}}
\put(7,7){\line(1,0){2}}
\put(0.5,7){\makebox(0,0)[r]{$x_{1}$}}
\put(0.5,3){\makebox(0,0)[r]{$x_{2}$}}
\put(9.5,7){\makebox(0,0)[l]{$\sim x_{1}$}}
\put(9.5,3){\makebox(0,0)[l]{$x_{2}.$}}
\end{picture}
\end{center}
This gate realizes Luckasiewicz negation for $x_{1}$:
$LN_{2}|x_1 \otimes x_2]=|(\sim x_1)\otimes x_2]$ iff $x_{2}=0 $.
If $x_{2}=0$, then the two-ququat gate must be following
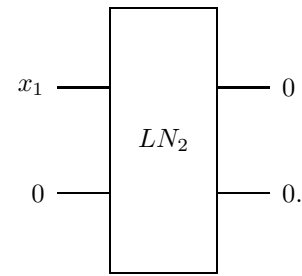
\begin{center}
\begin{picture}(10,10)(0,0)
\put(3,0){\framebox(4,10){$LN_{2}$}}
\put(1,3){\line(1,0){2}}
\put(1,7){\line(1,0){2}}
\put(7,3){\line(1,0){2}}
\put(7,7){\line(1,0){2}}
\put(0.5,7){\makebox(0,0)[r]{$x_{1}$}}
\put(0.5,3){\makebox(0,0)[r]{$0$}}
\put(9.5,7){\makebox(0,0)[l]{$0$}}
\put(9.5,3){\makebox(0,0)[l]{$0.$}}
\end{picture}
\end{center}

Let us write the unital quantum four-valued logic gate
which realizes Luckasiewicz negation in generalized
computational basis by
\[ \hat{\cal E}^{(LN_2)}=|00)(00|+\sum_{k=1,2,3}
\Bigl( |k 3)(k 0|+|k 2)(k 1|+\]
\[+|k 1)(k 2|+|k 0)(k 3|\Bigr). \]

By analogy to realization of Luckasiewicz negation we can derive
quantum gates with $T=0$ for classical gates $I_{0}(x)$,  $\overline{x}$,
$g_{1}$ and $g_{3}$.

\subsection{Unital quantum gates for two-arguments classical gates}

By analogy with Proposition 15 we can proof the following.

\bp
{\it The classical n-arguments 4-valued logic gate $g(x_1,...,x_n)$
can be realized as n-ququat unital quantum gate
if and only if $g(0,...,0)=0$, or $g(x_1,...,x_n)=const$.}
\ep

The two arguments nonconstant classical gate $g(x_{1},x_{2})$
can be realized by two-ququat linear unital quantum gate
$\hat{\cal E}$ if $g(0,0)=0$.


Two arguments nonconstant classical gates such that $g(0,0)\not=0$
can not be realized by two-ququat quantum gates $\hat{\cal E}$ with $T=0$.
These classical gates can be realized by three-ququats unital quantum gates.
Let us consider Sheffer function
$V_{4}(x_{1},x_{2})=max(x_{1},x_{2})+1(mod4)$.
If $x_{3}\not=0$ and $x_{1}, x_{2}, x_{3} \in \{0,1,2,3\}$, then
we can define unital quantum Sheffer gate by

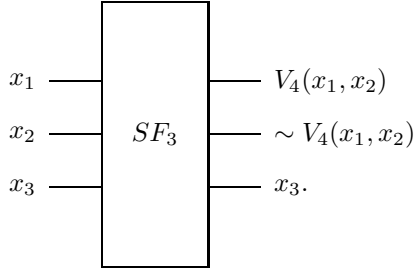
\begin{center}
\begin{picture}(10,10)(0,0)
\put(3,0){\framebox(4,10){$SF_{3}$}}
\put(1,3){\line(1,0){2}}
\put(1,5){\line(1,0){2}}
\put(1,7){\line(1,0){2}}
\put(7,3){\line(1,0){2}}
\put(7,5){\line(1,0){2}}
\put(7,7){\line(1,0){2}}
\put(0.5,7){\makebox(0,0)[r]{$x_{1}$}}
\put(0.5,5){\makebox(0,0)[r]{$x_{2}$}}
\put(0.5,3){\makebox(0,0)[r]{$x_{3}$}}
\put(9.5,7){\makebox(0,0)[l]{$V_{4}(x_{1},x_{2}) $}}
\put(9.5,5){\makebox(0,0)[l]{$\sim V_{4}(x_{1},x_{2}) $}}
\put(9.5,3){\makebox(0,0)[l]{$x_{3}.$}}
\end{picture}
\end{center}

If $x_{3}=0$ then we must have for unital quantum Sheffer gate

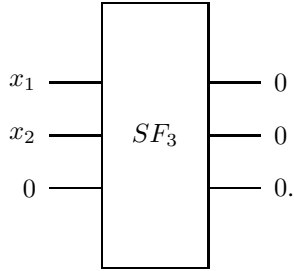
\begin{center}
\begin{picture}(10,10)(0,0)
\put(3,0){\framebox(4,10){$SF_{3}$}}
\put(1,3){\line(1,0){2}}
\put(1,5){\line(1,0){2}}
\put(1,7){\line(1,0){2}}
\put(7,3){\line(1,0){2}}
\put(7,5){\line(1,0){2}}
\put(7,7){\line(1,0){2}}
\put(0.5,7){\makebox(0,0)[r]{$x_1$}}
\put(0.5,5){\makebox(0,0)[r]{$x_2$}}
\put(0.5,3){\makebox(0,0)[r]{$0$}}
\put(9.5,7){\makebox(0,0)[l]{$0$}}
\put(9.5,5){\makebox(0,0)[l]{$0$}}
\put(9.5,3){\makebox(0,0)[l]{$0$.}}
\end{picture}
\end{center}

Let us write the unital quantum gate
which realizes Sheffer function in generalized
computational basis by
\[ \hat{\cal E}^{(SF_3)}=\sum^3_{\mu_1, \mu_2=0}|000)(\mu_1 \mu_20|+\]
\[ +\sum^3_{\mu_1, \mu_2=0}
 |V_4(\mu_1,\mu_2),\sim V_4(\mu_1,\mu_2),\mu_3)(\mu_1 \mu_2 \mu_3|. \]


\section{Universal set of quantum four-valued logic gates }



The condition for performing arbitrary unitary operations to
realize a quantum computation by dynamics of a closed quantum
system is well understood  \cite{Bar,DBE,DV,LL}. Using a universal
gate set, a quantum computer may realize the time sequence of
operations corresponding to any unitary dynamics. Deutsch, Barenco
and Ekert \cite{DBE}, DiVincenzo \cite{DV} and Lloyd \cite{LL}
showed that almost any two-qubits quantum gate is universal. It is
known \cite{Bar,DBE,DV,LL} that a set of quantum gates that
consists of all one-qubit gates and the two-qubits exclusive-or
(XOR) gate is universal in the sense that all unitary operations
on arbitrary many qubits can be expressed as compositions of these
gates. Recently in the paper \cite{BB} was considered universality
for n-qudits quantum gates.

The same is not true for the general quantum operations
(superoperators) corresponding to the dynamics of open quantum systems.
In the paper \cite{Bac} single qubit
open quantum system with Markovian dynamics was considered and
the resources needed for universality of general quantum
operations was studied.
An analysis of completely-positive trace-preserving superoperators on
single qubit density matrices was realized in papers \cite{FA,KR,RSW}.

Let us study universality for general quantum four-valued logic gates.
A set of quantum four-valued logic gates is universal iff all
quantum gates on arbitrary many ququats can be expressed
as compositions of these gates.
A set of quantum four-valued logic gates is universal iff all unitary
two-valued logic gates and general quantum operations
can be represented by compositions of these gates.
Single ququat gates cannot map two initially
un-entangled ququats into an entangled state.
Therefore the single ququat gates or set of single ququats gates
are not universal gates.
Quantum gates which are realization of classical gates
cannot be universal by definition, since these gates evolve
generalized computational states to generalized computational states
and never to the superposition of them.

Let us consider linear completely positive
trace-decreasing superoperator $\hat{\cal E}$.
This superoperator can be represented in the form
\begin{equation} \label{ELR} \hat{\cal E}=\sum^{m}_{j=1} \hat L_{A_{j}}
\hat R_{A^{\dagger}_{j}} \ , \end{equation}
where $\hat L_{A}$ and $\hat R_{A}$ are left and right multiplication
superoperators on $\overline{\cal H}^{(n)}$ defined by
$\hat L_{A}|B)=|AB)$, $\hat R_{A}|B)=|BA)$.

The n-ququats linear gate $\hat{\cal E}$ is completely positive
trace-preserving superoperator such that the gate matrix is an
element of Lie group $TGL(4^{n}-1,\mathbb{R})$. In general case, the
n-ququats nonlinear gate $\hat{\cal N}$ is defined by completely
positive trace-decreasing linear superoperator $\hat{\cal E}$ such
that the gate matrix is an element of Lie group $GL(4^{n},{\bf
R})$. The condition of completely positivity leads to difficult
inequalities for gate matrix elements \cite{Choi,FA,KR,RSW}. In
order to satisfy condition of completely positivity we use the
representation (\ref{ELR}). {\it To find the universal set of
completely positive (linear or nonlinear) gates $\hat{\cal E}$ we
consider the universal set of the superoperators $\hat L_{A_j}$
and $\hat R_{A^{\dagger}_j}$.} The matrices of these superoperators are
connected by complex conjugation. Obviously, the universal set of
superoperators $\hat L_{A}$ defines a universal set of completely
positive superoperators $\hat{\cal E}$ of the quantum gates. The
trace-preserving condition for linear superoperator (\ref{ELR}) is
equivalent to the requirement for gate matrix ${\cal E} \in
TGL(4^{n}-1,\mathbb{R})$, i.e. ${\cal E}_{0 \mu}=\delta_{0 \mu}$. The
trace-decreasing condition can be satisfied by inequality of the
following proposition.

\bp {\it If the matrix elements ${\cal E}_{\mu \nu}$ of a
superoperator $\hat{\cal E}$ is satisfied the inequality
\begin{equation} \label{ieq1}
\sum^{N-1}_{\mu=0} ({\cal E}_{0 \mu})^{2} \le 1 , \end{equation}
then  $\hat{\cal E}$ is a trace-decreasing superoperator.}
\ep

\noindent {\bf Proof.}
Using Schwarz inequality
\[ \Bigl( \sum^{N-1}_{\mu=0} {\cal E}_{0 \mu} \rho_{\mu} \Bigr)^{2}\le
\sum^{N-1}_{\mu=0} ({\cal E}_{0 \mu})^{2}
\sum^{N-1}_{\nu=0} (\rho_{\nu})^{2}  \]
and the property of density matrix
\[ Tr \rho^{2}=(\rho|\rho)=\sum^{N-1}_{\nu=0} (\rho_{\nu})^{2} \le 1, \]
we have
\[ |Tr\hat{\cal E}(\rho)|^{2}= |(0|\hat{\cal E}|\rho)|^{2}=
\Bigl(\sum^{N-1}_{\mu=0} {\cal E}_{0 \mu} \rho_{\mu} \Bigr)^{2}
\le \sum^{N-1}_{\mu=0} ({\cal E}_{0 \mu})^{2}. \]
Using (\ref{ieq1}), we get $|Tr \hat{\cal E}(\rho)| \le 1$.
Since $\hat{\cal E}$ is completely positive (or positive)
superoperator ($\hat{\cal E}(\rho) \ge 0$), it follows that
\[ 0 \le Tr \hat{\cal E}(\rho) \le 1, \]
i.e. $\hat{\cal E}$ is trace-decreasing superoperator.

Let the superoperators $\hat L_{A}$ and $\hat R_{A^{\dagger}}$
be called pseudo-gates. These superoperators can be represented by
\[ \hat L_A=  \sum^{N-1}_{\mu=0}\sum^{N-1}_{\nu=0}
L^{(A)}_{\mu \nu} |\mu)(\nu|, \quad
\hat R_{A^{\dagger}}= \sum^{N-1}_{\mu=0}  \sum^{N-1}_{\nu=0}
R^{(A^{\dagger})}_{\mu \nu} |\mu)(\nu|. \]

\bp
{\it The matrix of the completely positive superoperator (\ref{ELR})
can be represented by}
\begin{equation} \label{ELR-M} {\cal E}_{\mu \nu}
=\sum^{m}_{j=1} \sum^{N-1}_{\alpha=0} L^{(jA)}_{\mu \alpha}
R^{(jA^{\dagger})}_{\alpha \nu} \ . \end{equation}
\ep

\noindent {\bf Proof.}
Let us write the matrix ${\cal E}_{\mu \nu}$ by matrices of
superoperators $\hat L_{A_j}$ and $\hat R_{A_j}$.
\[ {\cal E}_{\mu \nu}=(\mu|\hat{\cal E} |\nu)=
\sum^{m}_{j=1} (\mu| \hat L_{A_j}
\hat R_{A^{\dagger}_j} |\nu)=\]
\[=\sum^{m}_{j=1} \sum^{N-1}_{\alpha=0} (\mu| \hat L_{A_j}|\alpha)
(\alpha|\hat R_{A^{\dagger}_j} |\nu)=
\sum^{m}_{j=1} \sum^{N-1}_{\alpha=0}  L^{(jA)}_{\mu \alpha}
R^{(jA^{\dagger})}_{\alpha \nu}. \]
Finally, we obtain (\ref{ELR-M}),
where
\[ L^{(jA)}_{\mu \alpha}=(\mu|\hat L_{A}|\alpha)=
\frac{1}{2^n} Tr\Bigl(\sigma_{\mu} \hat L_{A} \sigma_{\alpha} \Bigr)=\]
\[=\frac{1}{2^n} Tr\Bigl(\sigma_{\mu} A \sigma_{\alpha} \Bigr)=
\frac{1}{2^n} Tr\Bigl(\sigma_{\alpha} \sigma_{\mu} A  \Bigr), \]
\[ R^{(jA^{\dagger})}_{\alpha \nu}=(\alpha|\hat R_{A^{\dagger}}|\nu)=
\frac{1}{2^n} Tr\Bigl(\sigma_{\alpha} \hat R_{A^{\dagger}}
\sigma_{\nu} \Bigr)=\]
\[=\frac{1}{2^n}
Tr\Bigl(\sigma_{\alpha} \sigma_{\nu} A^{\dagger} \Bigr)
=\frac{1}{2^n}
Tr\Bigl(A^{\dagger}\sigma_{\alpha} \sigma_{\nu} \Bigr). \]
The matrix elements can be rewritten in the form
\begin{equation} \label{LRmat} L^{(jA)}_{\mu \alpha}=\frac{1}{2^n}
(\sigma_{\mu} \sigma_{\alpha}| A) \ , \quad
R^{(jA^{\dagger})}_{\alpha \nu}=
\frac{1}{2^n}(A|\sigma_{\alpha} \sigma_\nu). \end{equation}

\noindent{\bf Example.}
Let us consider the single ququat pseudo-gate $\hat L_{A}$.
The elements of pseudo-gate matrix $L^{(A)}$ are defined by
\[ L^{(A)}_{\mu \nu}=\frac{1}{2}Tr(\sigma_{\mu} A \sigma_{\nu}). \]
Let us denote
\[ a_{\mu}=\frac{1}{2}Tr(\sigma_{\mu} A). \]
Using
\[ L^{(A)}_{kl}=\frac{1}{2}Tr(\sigma_{l} \sigma_{k} A)=
\frac{1}{2}\delta_{kl}TrA +
\frac{i}{2}\varepsilon_{lkm} Tr(\sigma_{m} A), \]
where $k,l,m=1,2,3$, we get
\[ \hat L_{A}=\sum^{3}_{\mu=0} a_{0} |\mu)(\mu|+
\sum^{3}_{k=0} a_{k} \Bigl( |0)(k|+ |k)(0|\Bigr)+ \]
\[ +ia_{1}\Bigl( |3)(2|-|2)(3| \Bigr)+
ia_{2}\Bigl( |1)(3|-|3)(1| \Bigr)+\]
\[+ia_{3}\Bigl( |2)(1|-|1)(2| \Bigr). \]
The pseudo-gate matrix is
\[ L^{(A)}_{\mu \nu}= \delta_{\mu \nu}Tr A+
\sum^{3}_{m=1}\Bigl( \delta_{\mu 0}\delta_{\nu m} +
\delta_{\mu m}\delta_{\nu 0}\Bigr) Tr(\sigma_{m}A)+ \]
\[+i\sum^{3}_{m=1} \delta_{\mu k}\delta_{\nu l}
\varepsilon_{lkm}  Tr(\sigma_{m}A), \]
i.e.
$$L^{(A)}=\left(
\begin{array}{cccc}
a_{0}&a_{1}  &a_{2}  &a_{3}\\
a_{1}&a_{0}  &-ia_{3}&ia_{2}\\
a_{2}&ia_{3} &a_{0}  &-ia_{1}\\
a_{3}&-ia_{2}&ia_{1} &a_{0}
\end{array}
\right).$$

Let us consider properties of the matrix elements
$L^{(jA)}_{\mu \alpha}$ and $R^{(jA^{\dagger})}_{\mu \alpha}$.

\bp
{\it The matrices
$L^{(jA)}_{\mu \alpha}$ and $R^{(jA^{\dagger})}_{\mu \alpha}$
are complex $4^{n}\times 4^{n}$ matrices and their elements
are connected by complex conjugation: }
\[ (L^{(jA)}_{\mu \alpha})^{*}=R^{(jA^{\dagger})}_{\mu \alpha}. \]
\ep

\noindent {\bf Proof.}
Using complex conjugation of the matrix elements (\ref{LRmat}), we get
\[ (L^{(jA)}_{\mu \alpha})^{*}= \frac{1}{2^n}
(\sigma_{\mu} \sigma_{\alpha}| A)^{*}=
\frac{1}{2^n} (A|\sigma_{\mu} \sigma_{\alpha})=
R^{(jA^{\dagger})}_{\mu \alpha}. \]

We can write the gate matrix (\ref{ELR-M}) in the form
\[ {\cal E}_{\mu \nu}
=\sum^{m}_{j=1} \sum^{N-1}_{\alpha=0} L^{(jA)}_{\mu \alpha}
(L^{(jA)}_{\alpha \nu})^{*}. \]


\bp
{\it The matrices $L^{(jA)}_{\mu \alpha}$ and
$R^{(jA^{\dagger})}_{\mu \alpha}$ of the n-ququats
quantum gate (\ref{ELR}) are the elements
of Lie group $GL(4^{n},\mathbb{C})$.}
\ep

\noindent {\bf Proof.} The proof is trivial.


A two-ququats gate $\hat{\cal E}$ is called primitive \cite{BB} if
$\hat{\cal E}$ maps tensor product of single ququats
to tensor product of single ququats, i.e.
if $|\rho_{1})$ and $|\rho_{2})$ are ququats, then
we can find ququats $|\rho^{\prime}_{1})$ and $|\rho^{\prime}_{2})$
such that
\[ \hat{\cal E}|\rho_{1} \otimes\rho_{2})=
|\rho^{\prime}_{1} \otimes \rho^{\prime}_{2}). \]
The superoperator $ \hat{\cal E}$ is called imprimitive if
$\hat{\cal E}$ is not primitive.

It can be shown that almost every pseudo-gate that operates
on two or more ququats is universal pseudo-gate.

\bp
{\it The set of all single ququat pseudo-gates and
any imprimitive two-ququats pseudo-gate
are universal set of pseudo-gates.}
\ep


\noindent {\bf Proof.} This proposition can be proved by analogy
with \cite{DV,DBE,BB}. Let us consider some points of the proof.
Expressed in  group theory language, all n-ququats pseudo-gates
are elements of the Lie group $GL(4^n,\mathbb{C})$. Two-ququats
pseudo-gates $\hat L$ are elements of Lie group $GL(16,\mathbb{C})$.
The question of universality is the same as the question of what
set of superoperators $\hat L$ sufficient to generate $GL(16,\mathbb{C})$.
The group $GL(16,\mathbb{C})$ has $(16)^2=256$
independent one-parameter subgroups $GL_{\mu \nu}(16,\mathbb{C})$
of one-parameter pseudo-gates $\hat L^{(\mu \nu)}(t)$ such that
$\hat L^{(\mu \nu)}(t)=t|\mu)(\nu|$.
Infinitesimal generators of Lie group $GL(4^n,\mathbb{C})$ are defined by
\[ \hat H_{\mu \nu}=\Bigl(\frac{d}{dt} \hat L^{(\mu \nu)}(t) \Bigr)_{t=0}, \]
where $\mu,\nu=0,1,...,4^{n}-1$.
The generators $\hat H_{\mu \nu}$ of the one-parameter subgroup
$GL_{\mu \nu}(4^n,\mathbb{R})$ are superoperators of the form
$\hat H_{\mu \nu}=|\mu)(\nu|$ on $\overline{\cal H}^{(n)}$ 
which can be represented by
$4^n \times 4^n$ matrix $H_{\mu \nu}$ with elements
\[ (H_{\mu \nu})_{\alpha \beta}=\delta_{\alpha \mu} \delta_{\beta \nu}. \]
The set of superoperators $\hat H_{\mu \nu}$ is a basis 
(Weyl basis \cite{BR})
of Lie algebra $gl(16,\mathbb{R})$ such that
\[ [\hat H_{\mu \nu},\hat H_{\alpha \beta}]=
\delta_{\nu \alpha} \hat H_{\mu \beta}- \delta_{\mu \beta} \hat
H_{\nu \alpha}, \] 
where $\mu, \nu, \alpha, \beta =0,1,...,15.$
Any element $\hat H$ of the algebra
$gl(16,\mathbb{C})$ can be represented by
\[ \hat H=\sum^{15}_{\mu=0}\sum^{15}_{\nu=0}
h_{\mu \nu} \hat H_{\mu \nu}, \]
where $h_{\mu \nu}$ are complex coefficients.

As a basis of Lie algebra $gl(16,\mathbb{C})$ 
we can use $256$ linearly independent
self-adjoint superoperators
\[ H_{\alpha \alpha}=|\alpha)(\alpha|, \quad
H^r_{\alpha \beta}=|\alpha)(\beta|+|\beta)(\alpha|,\]
\[ H^i_{\alpha \beta}= -i\Bigl(
|\alpha)(\beta|-|\beta)(\alpha|\Bigr). \]
where $0\le \alpha \le \beta \le 15$.
The matrices of these generators is
Hermitian $16 \times 16$ matrices.
The matrix elements of 256 Hermitian $16 \times 16$ matrices
$H_{\alpha \alpha}$,  $H^r_{\alpha \beta}$ and $H^{i}_{\alpha \beta}$
are defined by
\[ (H_{\alpha \alpha})_{\mu \nu}=
\delta_{\mu \alpha} \delta_{\nu \alpha} \ ,
\quad (H^r_{\alpha \beta})_{\mu \nu}=
\delta_{\mu \alpha} \delta_{\nu \beta}
+\delta_{\mu \beta} \delta_{\nu \alpha}, \]
\[ (H^i_{\alpha \beta})_{\mu \nu}=
-i(\delta_{\mu \alpha} \delta_{\nu \beta}
-\delta_{\mu \beta} \delta_{\nu \alpha}). \]
For any Hermitian generators $\hat H$ exists
one-parameter pseudo-gates $\hat L(t)$ 
which can be represented in the form 
$\hat L(t)=exp \ it \hat H$ such that 
$\hat L^{\dagger}(t)\hat L(t)=\hat I$.

Let us write main operations which allow to derive
new pseudo-gates $\hat L$ from a set of pseudo-gates.\\
1) We introduce general SWAP (twist)
pseudo-gate $\hat T^{(SW)}$.
A new pseudo-gate $\hat L^{(SW)}$ defined by
$\hat L^{(SW)}=\hat T^{(SW)} \hat L \hat T^{(SW)}$
is obtained directly from $\hat L$ by exchanging two ququats.\\
2) Any superoperator $\hat L$ on $\overline{\cal H}^{(2)}$
generated by the commutator
$i[\hat H_{\mu\nu}, \hat H_{\alpha \beta}]$ can be obtained
from $\hat L_{\mu\nu}(t)=exp \ it\hat H_{\mu\nu}$
and $\hat L_{\alpha \beta}(t)=exp \ it\hat H_{\alpha \beta}$ because
\[ exp \ t \ [\hat H_{\mu\nu},\hat H_{\alpha \beta}]=\]
\[=\lim_{n \rightarrow \infty} \Bigl( \hat L_{\alpha \beta}(-t_n)
\hat L_{\mu\nu}(t_n) \hat L_{\alpha \beta}(t_n) \hat
L_{\mu\nu}(-t_n)\Bigr)^n, \]
where $t_n=1/\sqrt{n}$.
Thus we can use the commutator
$i[\hat H_{\mu \nu}, \hat H_{\alpha \beta}]$
to generate pseudo-gates.\\
3) Every transformation $\hat L(a,b)=exp i\hat H(a,b)$
of $GL(16,\mathbb{C})$ generated by
superoperator $\hat H(a,b)=a\hat H_{\mu\nu}+b\hat H_{\alpha \beta}$, where
a and b is complex, can obtained from
 $\hat L_{\mu\nu}(t)=exp \ it\hat H_{\mu\nu}$
and $\hat L_{\alpha \beta}(t)=exp \ it\hat H_{\alpha \beta}$   by
 \[ exp \ i \hat H(a,b)=
\lim_{n \rightarrow \infty} \Bigl( \hat L_{\mu \nu}(\frac{a}{n})
\hat L_{\alpha \beta}(\frac{b}{n})\Bigr)^n. \]

For other details of the proof, see \cite{DV,DBE,BB}
and \cite{Bare,Bar,LL}.


\section{Quantum four-valued logic gates of order (n,m) }


In general case, a quantum gate is defined to be the
most general quantum operation \cite{AKN}:

\noindent {\bf Definition}
{\it Quantum gate $\hat G$ of order (n,m) is a positive
(completely positive) linear (nonlinear) trace preserving map from density
matrix operator $|\rho)$ on $n$-ququats to density matrix operator
$|\rho^{\prime})$ on $m$-ququats.}

In the generalized computational (operator) basis the gate
$\hat G$ of order (n,m) can be represented by
Let us rewrite formula by
\begin{equation} \label{G}
\hat G^{(n,m)}=\frac{1}{\sqrt{2^n 2^m}}\sum^M_{\mu=0} \sum^N_{\nu=0}
G^{(n,m)}_{\mu \nu} |\sigma_{\mu})(\sigma_{\nu}|. \end{equation}
where
\[ \mu=\mu_1 4^{m-1}+...+\mu_{m-1}4+\mu_m,  \]
\[ \nu=\nu_1 4^{n-1}+...+\nu_{n-1}4+\nu_n  \ . \]
For the gate matrices we use $N=4^n-1$ and $M=4^m-1$.

The matrix $G^{(n,m)}_{\mu, \nu}$ of linear gate is
a real $4^n\times 4^m$-matrix with
\[ G^{(n,m)}_{0, \nu}=\delta_{0\nu}. \]
In general case, linear gates
$\hat G^{(n,m)}$ of order $(n,m)$ have
$G^{(n,m)}_{\mu 0}\not=0$, i.e. this gate is not unital.
i.e.
$$G^{(n,m)}=\left(
\begin{array}{ccccc}
1&0&0&...&0\\
T_1&R_{11}&R_{12}&...&R_{1N}\\
T_2&R_{21}&R_{22}&...&R_{2N}\\
...&...&...& ...&...\\
T_M&R_{M1}&R_{M2}&...&R_{MN}
\end{array}
\right).$$

\noindent {\bf Theorem 10.} (Singular Valued Decomposition for Matrix)\\
{\it Any real $N\times M$ matrix $G$ can be written in the form
\[ G={\cal U}_M D_{NM} {\cal U}^{\small T}_N, \]
where\\
${\cal U}_M$ is an orthogonal $M\times M$ matrix.\\
${\cal U}_N$ is an orthogonal $N\times N$ matrix.\\
$D_{NM}$ is diagonal $N\times M$ matrix such that
\[ D_{NM}=diag(\lambda_1,...,\lambda_p) \ , \quad p=min\{N,M\}, \]
where $\lambda_1 \ge \lambda_2 \ge ... \ge \lambda_p \ge 0$.
}

\noindent {\bf Proof.} This theorem is proved in \cite{EY,Lan,Sc,Gant}.


Let us consider the unital gates with $T=0$ defined by
\begin{equation} \label{RR} \hat G^{(n,m)}=\frac{1}{\sqrt{2^n 2^m}}
\Bigl(|0)(0|+ \sum^M_{\mu=1} \sum^N_{\nu=1}
G^{(n,m)}_{\mu, \nu} |\sigma_{\mu})(\sigma_{\nu}| \Bigr) \ . \end{equation}

\noindent {\bf Theorem 12.} (Singular Valued Decomposition for Gates)\\
{\it Any unital linear gate $\hat G^{(n,m)}$ of order $(n,m)$
defined by (\ref{RR})
can be represented by
\[ \hat G^{(n,m)}=
\hat{\cal U}^{(m,m)} \ \hat D^{(n,m)} \ \hat {\cal U}^{(n,n)}, \]
where\\
$\hat{\cal U}^{(m,m)}$ is an orthogonal quantum gate of order $(m,m)$.\\
$\hat{\cal U}^{(n,n)}$ is an orthogonal quantum gate of order $(n,n)$.\\
$\hat D^{(n,m)}$ is a diagonal quantum gate of order $(n,m)$, such that
\begin{equation} \label{Dnm} \hat D^{(n,m)}=\frac{1}{\sqrt{2^n 2^m}}
\Bigl( |0)(0|+\sum^p_{\mu=1} \lambda_{\mu}
|\sigma_{\mu})(\sigma_{\mu}| \Bigr),\end{equation}
where $p=min\{N,M\}$ and $\lambda_{\mu} \ge 0$.
}

\noindent {\bf Proof.} The proof of this theorem can be easy realized
in matrix representation by using theorem 10.

In general case, we have the following theorem.

\noindent {\bf Theorem 13.} (Singular Valued Decomposition for Gates)\\
{\it Any linear gate $\hat G^{(n,m)}$ of order $(n,m)$
can be represented by
\[ \hat G^{(n,m)}= \hat T^{(m,m)}\hat{\cal U}^{(m,m)} \
\hat D^{(n,m)} \ \hat{\cal U}^{(n,n)}, \]
where\\
$\hat{\cal U}^{(m,m)}$ is an
orthogonal quantum gate of order $(m,m)$.\\
$\hat{\cal U}^{(n,n)}$ is an
orthogonal quantum gate of order $(n,n)$.\\
$\hat D^{(n,m)}$ is a diagonal quantum gate (\ref{Dnm})
of order $(n,m)$.\\
$\hat T^{(m,m)}$ is a translation quantum gate
of order $(m,m)$:
\[\hat T^{(m,m)}=\frac{1}{\sqrt{2^n 2^m}}
\Bigl( |0)(0|+\sum^p_{\mu=1} |\sigma_{\mu})(\sigma_{\mu}|+
\sum^M_{\mu=0}
 T_{\mu} |\sigma_{\mu})(0| \Bigr), \]
where $p=min\{N,M\}$ and $\lambda_k \ge 0$.
}

\noindent {\bf Proof.} The proof of this theorem can be easy realized
in matrix representation by using theorem 10.

Note that, any n-arguments classical gate $g(\nu_{1},...,\nu_{n})$
can be realized as linear trace-preserving quantum
four-valued logic gate $\hat G^{(n,1)}$ of order $(n,1)$ by
\[ \hat G^{(n,1)}=|0)(0...0|+
\sum_{\nu_1...\nu_n \not=0...0}
|g(\nu_1,...,\nu_n))(\nu_1,...,\nu_n|+ \]
\[+ (1-\delta_{0g(0,...,0) }) |g(0,...,0))(0...0|-\]
\[- (1-\delta_{0g(0,...,0) })
\sum^{N-1}_{\mu=0} \sum_{\nu_1 ... \nu_n}
(1-\delta_{\mu g(\nu_1 ,...,\nu_n)})
|\mu)(\nu_1...\nu_n|. \]
In general case, this quantum gate is not unital gate.





\end{document}